\documentclass[10pt,twocolumn,showpacs,preprintnumbers,amsmath,amssymb,aps,prd,nofootinbib,superscriptaddress,longbibliography]{revtex4-2}
\usepackage{wasysym}
\usepackage[utf8]{inputenc}
\usepackage[english]{babel}
\usepackage{epsfig}
\usepackage{subfigure}
\usepackage{bm}
\usepackage{amsfonts}
\usepackage{dcolumn}
\usepackage{hyperref}

\hypersetup{colorlinks=true, linkcolor=blue, citecolor=green}

\usepackage{cleveref}
\usepackage{dcolumn}
\usepackage{bm}
\usepackage{ifpdf}
\usepackage{bm}
\usepackage{xcolor,color,graphicx,graphics}
\usepackage[OT1]{fontenc}
\usepackage{latexsym,amssymb,amsmath,amsfonts}
\usepackage{makeidx}
\usepackage{epstopdf}
\usepackage{mathrsfs}
\hypersetup{colorlinks=true, linkcolor=blue, citecolor=green}
\usepackage{enumerate}
 \usepackage{multirow}

 \usepackage{tikz}

\newcommand{\orcidicon}{%
	\begin{tikzpicture}
	\draw[lime, fill=lime] (0,0) 
		circle [radius=0.16] 
		node[white] {{\fontfamily{qag}\selectfont \tiny ID}};
	\draw[white, fill=white] (-0.0625,0.095) 
		circle [radius=0.007];
	\end{tikzpicture}	\hspace{-2mm}
}

\newcommand\orcidEdnaldo{{\href{https://orcid.org/0000-0001-7230-3666}{\orcidicon}}}
\newcommand\orcidFrancisco{{\href{https://orcid.org/0000-0002-9388-8373}{\orcidicon}}}
\newcommand\orcidManuel{{\href{https://orcid.org/0000-0001-8586-0285}{\orcidicon}}}
\newcommand\orcidTarciso{{\href{https://orcid.org/0009-0007-0450-2672}{\orcidicon}}}
\newcommand\orcidHenrique{{\href{https://orcid.org/0000-0001-7565-4277}{\orcidicon}}}
\newcommand\orcidLuis{{\href{https://orcid.org/0009-0009-4322-6484}{\orcidicon}}}
\newcommand\orcidJorde{{\href{https://orcid.org/0009-0001-3344-2986}{\orcidicon}}}
\newcommand\orcidDiego{{\href{https://orcid.org/0000-0003-3984-9864}{\orcidicon}}}

\begin{document}

\title{Accretion dynamics of thin and thick disks in charged Kalb-Ramond black holes}


\author{Ednaldo L. B. Junior\orcidEdnaldo\!\!} \email{ednaldobarrosjr@gmail.com}
\affiliation{Faculdade de F\'{i}sica, Universidade Federal do Pará, Campus Universitário de Tucuruí, CEP: 68464-000, Tucuruí, Pará, Brazil}
\affiliation{Programa de P\'{o}s-Gradua\c{c}\~{a}o em F\'{i}sica, Universidade Federal do Sul e Sudeste do Par\'{a}, 68500-000, Marab\'{a}, Par\'{a}, Brazil}

	\author{Jos\'{e} Tarciso S. S. Junior\orcidTarciso\!\!}
 \email{tarcisojunior17@gmail.com}
\affiliation{Faculdade de F\'{\i}sica, Programa de P\'{o}s-Gradua\c{c}\~{a}o em 
F\'isica, Universidade Federal do 
 Par\'{a},  66075-110, Bel\'{e}m, Par\'{a}, Brazil}

	\author{Francisco S. N. Lobo\orcidFrancisco\!\!} \email{fslobo@ciencias.ulisboa.pt}
\affiliation{Instituto de Astrof\'{i}sica e Ci\^{e}ncias do Espa\c{c}o, Faculdade de Ci\^{e}ncias da Universidade de Lisboa, Edifício C8, Campo Grande, P-1749-016 Lisbon, Portugal}
\affiliation{Departamento de F\'{i}sica, Faculdade de Ci\^{e}ncias da Universidade de Lisboa, Edif\'{i}cio C8, Campo Grande, P-1749-016 Lisbon, Portugal}

	\author{Jorde A. A. Ramos\orcidJorde\!\!}
 \email{jordealves@ufpa.br}
\affiliation{Faculdade de F\'{\i}sica, Programa de P\'{o}s-Gradua\c{c}\~{a}o em 
F\'isica, Universidade Federal do 
 Par\'{a},  66075-110, Bel\'{e}m, Par\'{a}, Brazil}
 
	\author{Manuel E. Rodrigues\orcidManuel\!\!}
	\email{esialg@gmail.com}
	\affiliation{Faculdade de F\'{\i}sica, Programa de P\'{o}s-Gradua\c{c}\~{a}o em 
F\'isica, Universidade Federal do 
 Par\'{a},  66075-110, Bel\'{e}m, Par\'{a}, Brazil}
\affiliation{Faculdade de Ci\^{e}ncias Exatas e Tecnologia, 
Universidade Federal do Par\'{a}\\
Campus Universit\'{a}rio de Abaetetuba, 68440-000, Abaetetuba, Par\'{a}, 
Brazil}


\author{Diego Rubiera-Garcia\orcidDiego\!\!} \email{drubiera@ucm.es}
\affiliation{Departamento de Física Teórica and IPARCOS, Universidad Complutense de Madrid, E-28040 Madrid, Spain}


\author{Luís F. Dias da Silva\orcidLuis\!\!} 
        \email{fc53497@alunos.fc.ul.pt}
\affiliation{Instituto de Astrof\'{i}sica e Ci\^{e}ncias do Espa\c{c}o, Faculdade de Ci\^{e}ncias da Universidade de Lisboa, Edifício C8, Campo Grande, P-1749-016 Lisbon, Portugal}


 \author{Henrique A. Vieira\orcidHenrique\!\!} \email{henriquefisica2017@gmail.com}
\affiliation{Faculdade de F\'{i}sica, Programa de P\'{o}s-Gradua\c{c}\~{a}o em F\'{i}sica, Universidade Federal do Par\'{a}, 66075-110, Bel\'{e}m, Par\'{a}, Brazil}

\date{\LaTeX-ed \today}
     

\begin{abstract}
In this work, we investigate the accretion properties of an electrically charged, static, spherically symmetric black hole in  Kalb-Ramond (KR) gravity, where a dimensionless parameter $l$ characterizes spontaneous Lorentz symmetry-breaking. We first constrain the free charge and KR parameter space, $(Q,l)$, by comparing the predicted shadow radius of this family of solutions with the bounds inferred from Event Horizon Telescope observations of Sagittarius A*. The inclusion of electric charge introduces a degeneracy between $Q$ and $l$, enlarging the region compatible with the observed shadow size and yielding the broad constraint $-0.3900 \leq l \leq 0.1898$, with part of the compatible parameter space extending into the naked-singularity sector. We then investigate relativistic accretion disks around the charged KR black hole within the above constraints. For geometrically thin disks, we compute the time averaged energy flux, temperature distribution, emission spectrum, time averaged torque, and mass-to-radiation conversion efficiency, finding that the radiative output increases for higher values of $l$, while lower values of $l$ suppress it. Furthermore, when coupled with the electric charge this effect is amplified. For geometrically thick, constant-angular momentum tori-shaped disks, Lorentz symmetry-breaking shifts the characteristic orbital radii, compresses the radial extent of equilibrium configurations, narrows the interval supporting finite equilibrium tori, and shifts the stationary-fluid topology toward open and unconfined configurations outside that interval. These results show how charge and spontaneous Lorentz symmetry-breaking jointly affect observational constraints and the structure and emission properties of black hole accretion flows, providing potential signatures of KR gravity in the strong-gravity regime.
\end{abstract}

\date{\today}

\maketitle

\def\HMS{{\scriptscriptstyle{\rm HMS}}}

\section{Introduction}

Over the past century, a series of observational breakthroughs have consistently reinforced the predictions of General Relativity (GR). Regarding compact objects, the first direct detection of gravitational waves by the LIGO collaboration in 2015, originating from a binary black hole merger, provided a striking confirmation of this prediction of Einstein’s theory \cite{LIGOScientific:2016aoc, LIGOScientific:2016vlm}. More recently, the Event Horizon Telescope (EHT) collaboration captured the first images of the supermassive object at the center of the galaxy M87 in 2019 \cite{EventHorizonTelescope:2019dse, EventHorizonTelescope:2019uob, EventHorizonTelescope:2019jan, EventHorizonTelescope:2019ths, EventHorizonTelescope:2019pgp, EventHorizonTelescope:2019ggy}, followed by the image of Sagittarius A* (Sgr A*) at the center of the Milky Way {\color{blue}\cite{EventHorizonTelescope:2022wkp, EventHorizonTelescope:2022apq, EventHorizonTelescope:2022wok, EventHorizonTelescope:2022exc, EventHorizonTelescope:2022urf, EventHorizonTelescope:2022xqj}}. Furthermore, the work  {\color{blue}\cite{Vagnozzi:2022moj}} systematically analyzed how EHT observations of Sgr A* place stringent constraints on black hole models that predict a shadow size significantly departing from that of a Schwarzschild black hole for a given mass. These observations have reinforced both the existence of black holes and the reliability of GR in describing gravity across the scales tested so far.

On the other hand, local phenomena play a fundamental role in GR, particularly in the accretion of matter onto black holes. 
Matter accretion was first described by Hermann Bondi \cite{bondi1952spherically, Armitage:2020owb} within a Newtonian framework, assuming a spherically symmetric, non-rotating flow. This model was later extended to GR by Frederick C. Michel \cite{Michel1972}, preserving the radial symmetry. Subsequent studies explored its stability \cite{Moncrief1980} and revealed that the nature of the accreting fluid is crucial, as exotic components such as phantom energy may even lead to black hole mass loss \cite{Babichev:2004yx, Babichev:2013vji, Jamil:2010skm}.
A more realistic description emerged with Igor Novikov and Kip Thorne  \cite{novikov1973astrophysics}, who introduced relativistic accretion disks with angular momentum, later formalized by Don Page and Thorne \cite{page1974disk}, establishing the standard and well-known thin disk Novikov-Thorne (NT) model.

Matter accretion onto black holes represents one of the most efficient mechanisms of energy release in the universe, playing a fundamental role in the understanding of X-ray binaries, active galactic nuclei, and gamma ray bursts \cite{Lahiri:2020sza, Chen:2024jsv}, in addition to the imaging of supermassive black holes \cite{Ricarte:2022gcl,Dhruv:2024igk}. As such, it has been the source of numerous investigations in the community \cite{Abramowicz:2011xu}. In particular, within the framework of relativistic hydrodynamics, a classical model describing such systems consists of geometrically thick, non-self-gravitating tori \cite{Font:2002bi, Rezzolla:2003re}. These models are based on barotropic perfect fluids orbiting the central object with constant specific angular momentum, satisfying the relativistic Euler equation. The disk structure is determined by equipotential surfaces and exhibits a key feature: the presence of a cusp at the inner edge of the disk. This cusp acts as a saddle point of the effective potential, allowing matter to flow toward the compact object through radial pressure gradients, without requiring angular momentum transport mechanisms such as viscosity. However, constant angular momentum disks are intrinsically prone to runaway instability \cite{Font:2002bi}. This instability arises from a dynamical feedback process in which mass accretion increases the black hole mass, modifying the gravitational field and shifting the cusp deeper into the disk. As a consequence, the mass transfer rate is amplified, potentially leading to the complete destruction of the disk on dynamical timescales, typically within only a few orbital periods. When other effects, such as black hole rotation, magnetic fields, and shear viscosity influenced by space-time curvature, are included, they may act as stabilizing mechanisms, suppressing or mitigating the instability growth \cite{Font:2002bi}. Moreover, accretion dynamics in alternative space-times, such as Reissner-Nordstr\"om (RN) naked singularities {\color{blue}\cite{PhysRevD.110.124030}} or black holes embedded in swirling universes {\color{blue}\cite{Chen:2024jsv}}, introduces new phenomenology, including zero-gravity spheres and matter outflows that challenge the predictions of purely stable models.

On the gravitational side, and despite GR successes, observations and theoretical consistency issues, such as the unavoidable existence of space-time singularities inside black holes and in the early Universe \cite{Senovilla:2014gza}, or the difficulty of reconciling GR and quantum mechanics \cite{Carlip:2001wq}, still allow for possible extensions of the GR framework in the strong-field regime, motivating the exploration of modified theories of gravity that may predict new compact objects and phenomena. One avenue to extend GR involves the violation of Lorentz invariance, a fundamental symmetry of GR that ensures the equivalence of local reference frames, at high-enough energies or curvature scales. This is a possibility admitted across a broad range of approaches, such as  string theory \cite{Kostelecky:1988zi, PhysRevLett.63.224, PhysRevD.40.1886, Colladay:1996iz}, loop quantum gravity \cite{Rovelli2004, Ashtekar:2004eh, Gambini:1998it, Ellis:1999uh}, non-commutative field theory \cite{Douglas:2001ba, Carroll:2001ws, Szabo:2001kg}, massive gravity \cite{deRham:2014zqa, Rham:2015mxa}, or Horava-Lifshitz gravity \cite{Horava:2009uw}. For a review on the theoretical and observational status of Lorentz invariance-violating (LIV) theories see e.g. \cite{Addazi:2021xuf}.

Lorentz symmetry-breaking can occur either explicitly, when the Lagrangian is not Lorentz invariant and introduces preferred frames \cite{Kostelecky:2003fs}, or spontaneously, when the Lagrangian remains invariant but a field acquires a vacuum expectation value that selects a preferred direction in space-time \cite{Kostelecky:1991ak, PhysRevLett.66.1811, Kostelecky:1995qk, Kostelecky:2000hz,Myers:2003fd, Bluhm:2008yt, Casana:2017jkc}. A particularly relevant example of this implementation is provided by the  antisymmetric Kalb--Ramond (KR) tensor field, originally introduced in Ref.~\cite{Kalb:1974yc}. In gravitational models with such an antisymmetric tensor background, spontaneous Lorentz symmetry-breaking can be realized when such a field acquires a nonzero vacuum expectation value and couples nonminimally to curvature \cite{Altschul:2009ae,Aashish:2019ykb}. Within this framework, in \cite{Duan:2023gng}, an exact static, spherically symmetric solution for electrically charged black holes was derived, corresponding to the KR field coupled to the Einstein-Hilbert action and an electromagnetic field \cite{Altschul:2009ae, Lessa:2019bgi}. As its uncharged counterpart, such a solution is characterized by the KR parameter $l$, in addition to the electric charge $Q$.

The main aim of this work is to perform a thorough study of accretion disk dynamics around the above electrically charged KR model. We shall first leverage the degeneracy of the parameter space ($Q,l$) to enlarge the allowed range of values of the KR parameter as compared to the uncharged case, the latter studied in \cite{Junior:2024ety}, using the inferred size of the shadow of Sgr A$^*$ by the EHT Collaboration. Within the obtained bounds, we shall consider both geometrically thin and thick disks, investigating different observational signatures and discussing how the spontaneous Lorentz symmetry-breaking KR parameter $l$ may affect  such signatures.

This work is organized as follows. In Sec.\,\ref{II}, we obtain new constraints on the LIV parameter $l$ using EHT observations, considering the charged black hole KR solution mentioned above. In Sec.\,\ref{III}, we study the properties of thin accretion discs for this solution, including the behavior of the energy flux, disk temperature, and luminosity.  In Sec.\,\ref{IV}, we investigate the equilibrium structure and stability of relativistic thick accretion discs in the KR framework, analyzing the behavior of equipotential surfaces, characteristic orbits, and the transition between confined and unconfined stationary equilibrium configurations. In particular, we show that the $l$ parameter compresses the stable disk region and anticipates the transition to unstable configurations. And finally, Sec.\,\ref{Sec:Conclusion} gathers the conclusion of our work.

\section{Constraints for the Lorentz violation parameter on charged KR black holes} \label{II}

\subsection{Space-time geometry}

The electrically charged solution of KR considered throughout this work, found in \cite{Duan:2023gng}, is conveniently written, in spherical coordinates $(t,r,\theta,\phi)$, as
\begin{equation}
ds^2 = -A(r)dt^2+B(r)dr^2+ C(r)(d\theta^2+\sin^2\theta d\phi^2)\,.\label{eq:generic_metric}
\end{equation}
with the metric functions\footnote{We adopt a Planck unit convention, setting $G=c=\hbar=k_B=1$.} 
\begin{eqnarray}
\label{eq:line_element}A(r)&=&\frac{1}{1-l}-\frac{2 M}{r}+\frac{Q^2}{(1-l)^2 r^2},\\
\label{eq:line_element2}B(r)&=&A^{-1}(r),\\
\label{eq:line_element3}C(r)&=&r^2,
\end{eqnarray}
where $M$ is the mass parameter, $Q$ is the electric charge, and $l$ is an additional dimensionless parameter that characterizes the spontaneous Lorentz symmetry-breaking. 

For the charged KR space-time \eqref{eq:generic_metric}-\eqref{eq:line_element3}, the horizon structure depends on the interplay between the parameters $Q$ and $l$. Throughout the black hole analysis we restrict to $1-l>0$. The transition between black-hole and naked-singularity configurations occurs at
\begin{equation}
Q^2=M^2(1-l)^3,
\end{equation}
for which the two horizons are degenerate. Thus, the black hole sector satisfies $Q^2\leq M^2(1-l)^3$, while $Q^2>M^2(1-l)^3$ corresponds to a naked singularity. If one restricts to $Q\geq0$, this is equivalently written as $Q\leq Q_c$, with $Q_c=M(1-l)^{3/2}$.
Although the absence of an event horizon does not preclude the existence of a central dark region in the corresponding images, see e.g. \cite{Rosa:2023qcv}, the distinction remains relevant in the present context, since the accretion-disk analysis performed in the following sections is restricted to black hole configurations.

\subsection{Constraints on $l$ from EHT shadow's radius inference}

In the uncharged case, $Q=0$, constraints on the Lorentz symmetry-breaking parameter can be obtained from orbital precession analyses, which yields the range \cite{Junior:2024ety}:
\begin{equation}\label{eq:KRconstraints}
-0.185022 \leq l \leq 0.0609.
\end{equation}
for observational consistency. The addition of an electric charge to this setting allows to broaden the range of compatibility with observations of the charged KR geometry above. To this end, we shall assess the compatibility of its predicted shadow radius with the observational  bounds of Sgr A*, as reported by the EHT Collaboration \cite{EventHorizonTelescope:2022urf}. The shadow is understood here as the central dark region resulting from those light rays that are captured by the black hole when a backwards ray-tracing procedure from the observer's screen towards the black hole is performed. It is bounded by a critical curve corresponding to the projection of the photon sphere, the latter defined as the surface of unstable bound null geodesics around the black hole \cite{Cunha:2018acu}. This shadow thus provides a direct probe of the underlying space-time geometry  and thus constitutes a useful observable to constrain any deviations from GR predictions\footnote{It should be pointed out that, in general, the shadow does not coincide with the central dark region in the image, the latter determined by a complex interplay between space-time geometry and accretion disk physics, see e.g. \cite{Gralla:2019xty,Chael:2021rjo}.}. For an asymptotically flat and spherically symmetric space-time, the shadow radius is given by \cite{perlick2022}
\begin{equation}
r_{\rm sh} = \sqrt{\frac{C(r_{\rm ph})}{A(r_{\rm ph})}},
\end{equation}
where $r_{\rm ph}$ denotes the radius of the photon sphere, obtained from the condition \cite{claudel2001}
\begin{equation} \label{eq:ps}
\frac{d}{dr}\left(\frac{C(r)}{A(r)}\right)\Big|_{r=r_{\rm ph}} = 0.
\end{equation}
In the case of the charged KR geometry considered in this work, the shadow radius  depends on the parameters $(M,Q,l)$, and can be written as
\begin{equation}
r_{\rm sh}=\frac{r_{\rm ph}}{\sqrt{1-\frac{2M(1-l)}{r_{ph}}+\frac{Q^2}{(1-l) r_{ph}^2}}}.
\end{equation}
where $r_{ph}$ can be explicitly written upon resolution of Eq.(\ref{eq:ps}) with the KR metric functions of Eqs.(\ref{eq:line_element})-(\ref{eq:line_element3}). For convenience, in our analysis all quantities shall be normalized with respect to the mass parameter $M$ (i.e., $Q$ is expressed in units of $M$). This effectively fixes the overall scale of the system, allowing us to freely explore the parameter space of $(Q,l)$, and confront the theoretical prediction for the shadow radius with the observational bounds inferred by the EHT collaboration for Sgr A*.

Since the shadow itself cannot be directly resolved due to the limited instrumental sensitivity to low-intensity photons, its size is instead inferred from the angular diameter of the surrounding bright emission ring, which acts as a proxy for the shadow's radius once appropriate calibration factors are taken into account (see \cite{EventHorizonTelescope:2022urf} for a detailed discussion). To implement this, the EHT collaboration relies on independent measurements of the mass-to-distance ratio $M/D$ of Sgr A*, obtained through orbital monitoring of the S-stars by the Keck Observatory \cite{Do:2019txf} and the Very Large Telescope Interferometer (VLTI) \cite{GRAVITY:2020gka}. Using these inputs, the EHT collaboration quantifies the fractional deviation $\delta$ between the inferred shadow diameter and the shadow diameter predicted  for a Schwarzschild black hole, whose angular shadow size is given by $\theta_{\rm sh,Sch}=6\sqrt{3}\,M/D$ \cite{EventHorizonTelescope:2022urf}. Such a fractional deviation is thus defined as 
\begin{equation}\label{eq:fractional_deviation}
    \delta = \frac{r_{sh}}{3\sqrt{3}M} - 1 \ .
\end{equation}
Assuming a gaussian distribution of the uncertainties, this quantity can be recast as bounds on the dimensionless shadow radius  as \cite{Vagnozzi:2022moj}
\begin{equation}
    4.55 \lesssim r_{s}/M \lesssim 5.22 \quad (1\sigma),
\end{equation}
and 
\begin{equation}
    4.21 \lesssim r_{s}/M \lesssim 5.56 \quad (2\sigma).
\end{equation}

\begin{figure}[t!]
    \centering
    \includegraphics[width=\columnwidth]{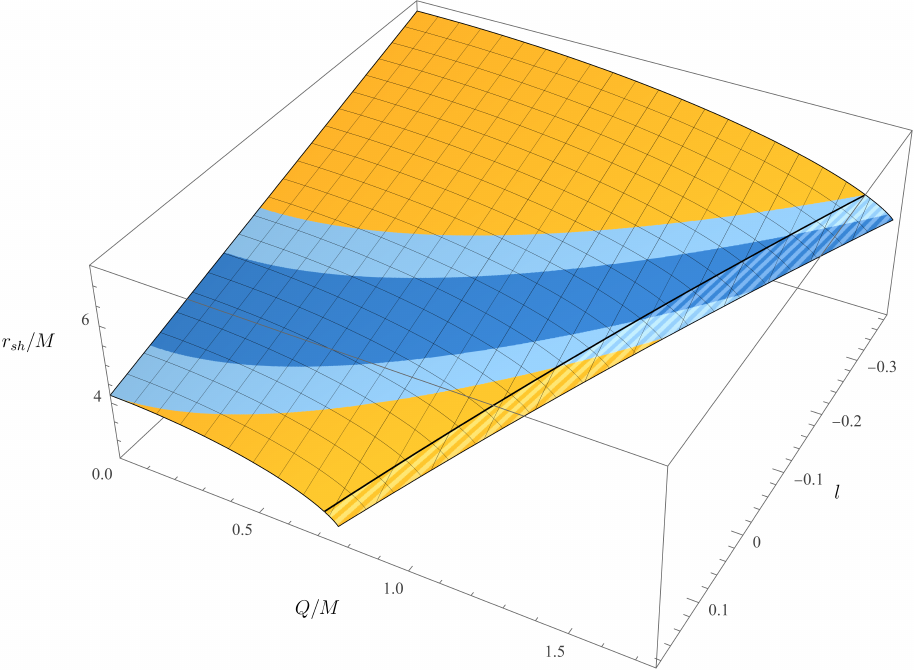}
    \caption{Surface corresponding to the theoretical shadow radius $r_{\rm sh}$ of the charged KR space-time \eqref{eq:line_element}-\eqref{eq:line_element3} as a function of the charge $Q$ and the Lorentz symmetry-breaking parameter $l$. The uniform colored surface corresponds to the black hole parameter domain. The striped surface corresponds to the naked singularity parameter domain. The solid black line separating the two regions corresponds to the phase boundary between black holes and horizonless configurations. The blue shaded areas represent the confidence intervals of Sgr A* shadow radius at 1$\sigma$ (dark blue) and at 2$\sigma$ (light blue).}
    \label{fig:shadowradius3D}
\end{figure}

For our charged KR solution, we numerically compute the  shadow radius $r_{\rm sh}$ across a range of values within the $(Q,l)$ parameter space and present the results in Fig.~\ref{fig:shadowradius3D}. While increasing each parameter $Q$ and $l$ individually tends to reduce the shadow size, the presence of a non-vanishing charge significantly enlarges the region of parameter space compatible with the EHT bounds. In particular, compared to the neutral Kalb–Ramond case ($Q/M=0$) \eqref{eq:KRconstraints}, the introduction of charge relaxes the constraints on $l$ to
\begin{equation}\label{eq:new_KRconstraints}
-0.3900 \leq l \leq 0.1898
\end{equation}
essentially allowing both larger and smaller values of this parameter to remain consistent with the observed shadow size of Sgr A*. This, in turn, allows a subset of naked singularity configurations of the charged KR field to remain compatible with the EHT shadow size bounds, for sufficiently negative values of $l$. Moreover, it draws attention to the degeneracy introduced by the two parameters, as multiple combinations of $(Q,l)$ predict the same theoretical shadow radius.

\begin{figure}[ht!]
    \centering
    \includegraphics[width=\columnwidth]{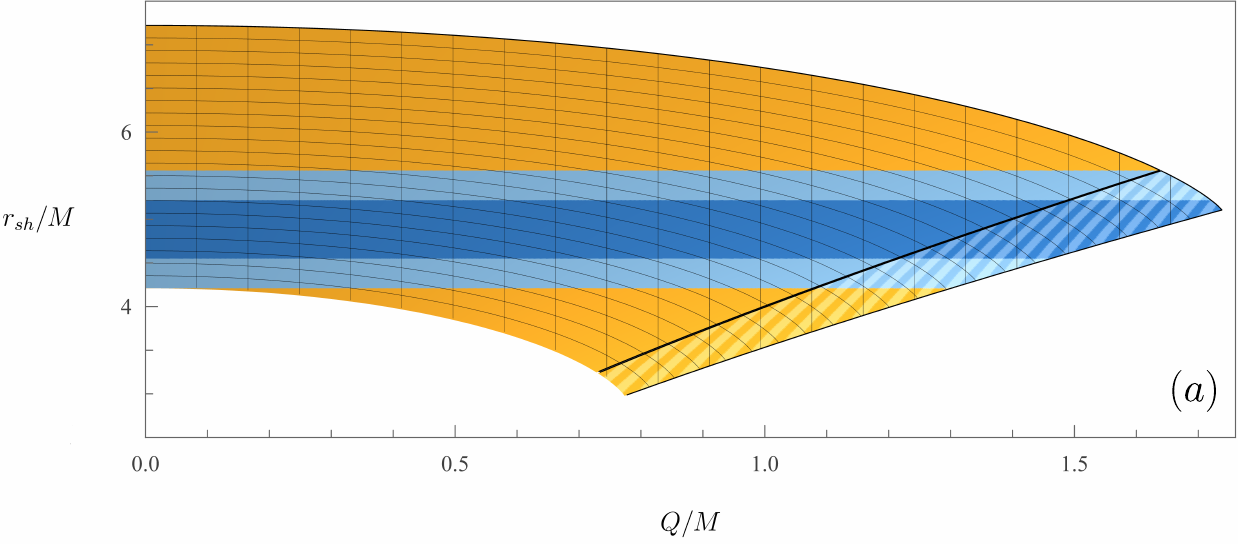}
    \includegraphics[width=\columnwidth]{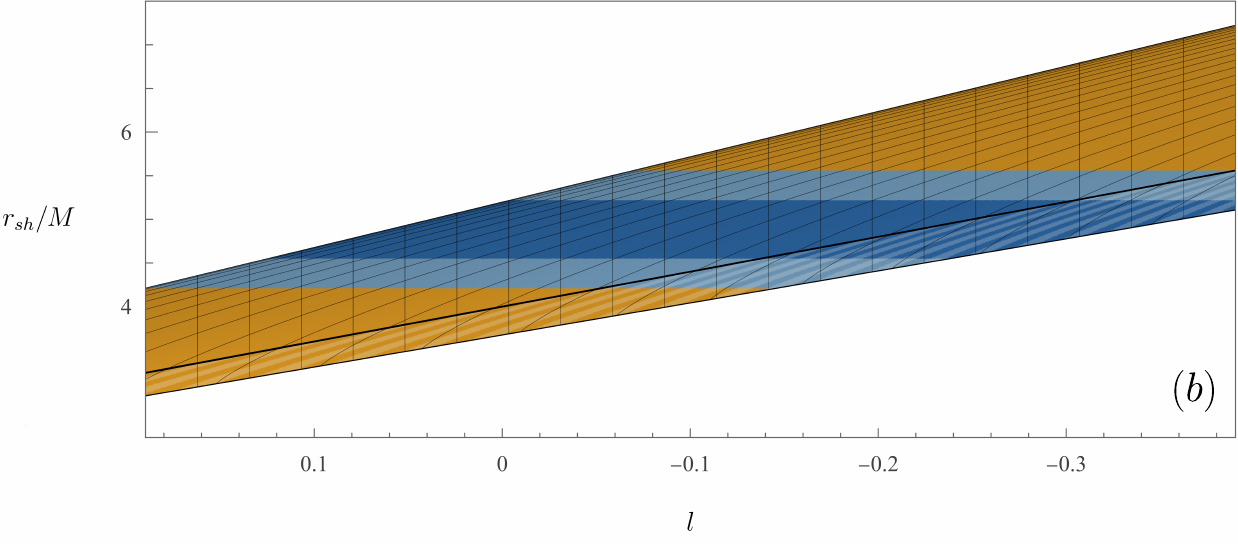}
    \includegraphics[width=\columnwidth]{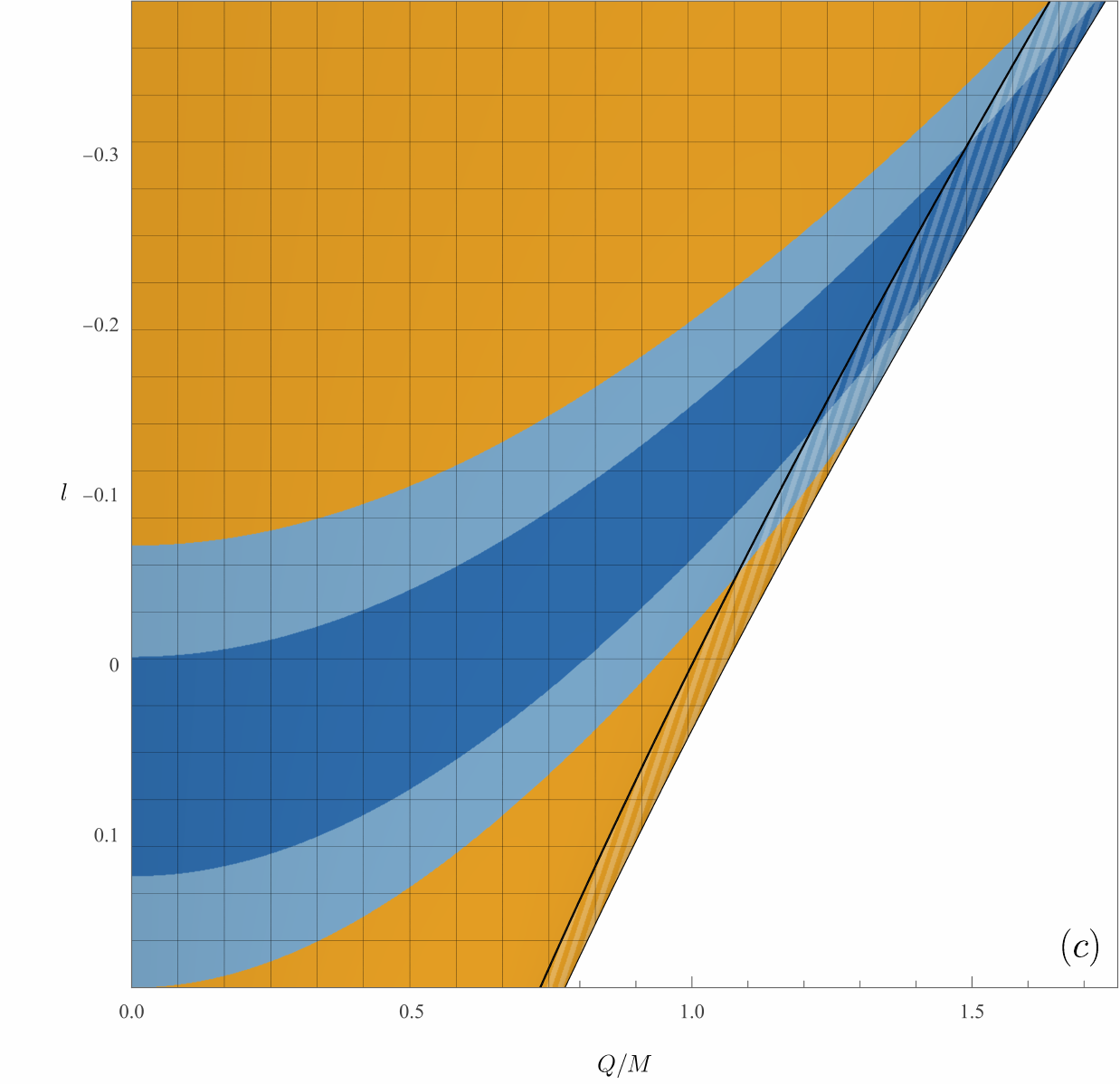}
    \caption{Two-dimensional projections of the plot from Fig.~\ref{fig:shadowradius3D}. The $(r_{\rm sh},Q)$, $(r_{\rm sh},l)$, and $(Q,l)$ planes correspond to the top $(a)$, middle $(b)$, and bottom $(c)$ sub-figures, respectively. The uniform colored surface corresponds to the black hole parameter domain. The striped surface corresponds to the naked singularity parameter domain. The solid black line separating the two regions corresponds to the phase boundary between black holes and horizonless configurations. The blue shaded areas represent the confidence intervals of Sgr A*'s shadow radius at 1$\sigma$ (dark blue) and at 2$\sigma$ (light blue).}
    \label{fig:shadowradius2D_abc}
\end{figure} 

While Fig. \ref{fig:shadowradius3D} provides a global view of the shadow-radius dependence on the parameters $(Q,l)$, its two-dimensional projections shown in Fig. \ref{fig:shadowradius2D_abc}, covering the $(r_{\rm sh},Q)$, $(r_{\rm sh},l)$, and $(Q,l)$ planes, respectively, allow the role of each parameter to be examined separately. The projection onto the $(r_{\rm sh}, Q)$ plane is shown in Fig.~\ref{fig:shadowradius2D_abc}$(a)$, corresponding to the family of shadow radius curves $r_{\rm sh}(Q; l)$ parametrized by $l$. As can be seen from this figure, for a fixed value of $l$, the shadow radius decreases monotonically with the charge parameter, reflecting the RN-like character of the geometry. Varying $l$ shifts these curves vertically, allowing different charge configurations to enter or leave the region compatible with the EHT bounds (blue bands). As a consequence, parameter values and configurations that would otherwise be excluded in the neutral KR case may remain consistent with current observational data.

The projection onto the $(r_{\rm sh}, l)$ plane is shown in Fig.~\ref{fig:shadowradius2D_abc}$(b)$. Analogously to the previous case, this projection corresponds to a family of shadow-radius curves, $r_{\rm sh}(l;Q)$, parametrized by the charge $Q$, whose upper boundary is defined by the neutral KR solution ($Q/M=0$). For a fixed value of $Q$, the shadow radius increases as $l$ decreases, while varying the charge shifts the corresponding curves vertically.  Consequently, each charge configuration is associated with a distinct interval of $l$ compatible with the EHT bounds. The broad constraint (\ref{eq:new_KRconstraints}) therefore emerges from the union of these admissible intervals, illustrating how the inclusion of charge relaxes the bounds obtained in \cite{Junior:2024ety}.

Lastly, Fig.~\ref{fig:shadowradius2D_abc}$(c)$, displays the projection onto the $(Q,l)$ plane, mapping the EHT constraints directly onto the parameter space itself. In this case, the blue shaded areas correspond to the portions of the  parameter space whose predicted shadow radii lie within the EHT bounds at $1\sigma$ (dark blue) and $2\sigma$ (light blue). This provides a clear visualization of the admissible parameter space, allowing one to immediately identify whether a space-time with a given set of $(Q,l)$ values remains compatible with current observational data. As with the previous cases, the black solid line denotes the critical charge separating black hole and naked singularity configurations, illustrating that both sectors contain solutions consistent with the EHT shadow-size constraints. A closer examination of this figure reveals that the broad constraint (\ref{eq:new_KRconstraints}) emerges from the union of the charge-dependent intervals compatible with the EHT bounds. As the charge increases, the corresponding admissible range of $l$ values gradually tightens and shifts towards more negative values; a behavior that exhibits a non-linear dependence on $Q$. This is consistent with the trends observed in the previous projections and further illustrates the interplay between the charge and Lorentz symmetry-breaking parameters in determining the shadow radius: the geometry's shadow radius becomes more sensitive to LIV the more charged the space-time is. The reason why the compatible parameter space of $l$ is shifted to more negative values as $Q$ increases is because their effects on the shadow radius balance out, namely, the increase in shadow radius due to negative values of $l$ compensates for the decrease in shadow radius caused by higher values of the charge $Q$.

\section{Thin accretion disks} \label{III}

We consider the relativistic description of a geometrically thin, optically thick, steady-state (i.e. disk height $H$ much smaller than characteristic radius $R$) accretion disk  located on the equatorial plane of the charged KR space-time characterized by Eqs.(\ref{eq:line_element})-(\ref{eq:line_element3}).  The accretion flow is assumed to be in hydrodynamical equilibrium, with negligible pressure gradient and vertical entropy gradient in the orbiting matter. Such equilibrium is maintained by efficient cooling mechanisms via radiation transport, which stabilizes its thin structure. In steady-state accretion disk models, the physical quantities describing the orbiting matter are averaged over a characteristic time scale, $\Delta t$, for a total period of the orbits, over the azimuthal angle $\Delta \phi = 2\pi$, and over the disk height $H$ \cite{novikov1973astrophysics,page1974disk}. For example, when averaged over $\Delta t, \Delta \phi$, and $H$, the baryons in the orbiting plasma move in nearly circular (Keplerian) orbits with angular velocity $\Omega=u^{\phi}/u^t$, having a specific energy $\widetilde{E}=-u_t$ and specific angular momentum $\widetilde{L}=u_{\phi}$. In this modeling, the mass accretion rate $\dot{M}_0$ is assumed to be a constant that does not change with time.

Considering a spherically symmetric space-time given by a line element \eqref{eq:generic_metric}, the conserved specific energy and angular momentum satisfy the equations
\begin{eqnarray}\label{eq:Conserved_EandL}
    \dot{t}=\frac{\widetilde{E}}{A(r)},
    \qquad
    \dot{\phi}=\frac{\widetilde{L}}{C(r)},
\end{eqnarray}
respectively, where an overdot denotes differentiation with respect to the proper time. The geodesic equation of motion for timelike particles may be expressed as
\begin{eqnarray}
   \dot{r}^2=\widetilde{E}^2-A(r)\left(1+\frac{\widetilde{L}^2}{C(r)}\right)\,,
\end{eqnarray}
which essentially takes the form of a one-dimensional radial effective-potential equation
\begin{eqnarray}\label{Veff}
    V_{\rm eff}(r)= A(r)\left(1+\frac{\tilde{L}^2}{C(r)}\right)\,,
\end{eqnarray}
where $V_{\rm eff}(r)$ denotes the effective potential. The conditions for circular geodesic motion, $V_{\rm eff}(r)=\tilde{E}^2$ and $V_{{\rm eff}, r} (r)=0 $, yield
\begin{eqnarray}
    \tilde{E} &=& \frac{A(r)}{\sqrt{A(r)-\Omega^2 C(r)}},\label{eq:specific_E}\\
    \tilde{L} &=& \frac{\Omega \, C(r)}{\sqrt{A(r)-\Omega^2 C(r)}},\label{eq:specific_L}\\
    \Omega &=& \sqrt{\frac{A'(r)}{C'(r)}},\label{eq:Omega}
\end{eqnarray}
where the prime notation denotes a derivative with respect to the radial coordinate $r$.

The physical parameters of the disk (e.g., flux, temperature, torque) can be expressed in terms of the angular velocity, specific energy, and specific angular momentum, which are the dynamical quantities presented in Eqs.(\ref{eq:specific_E}), (\ref{eq:specific_L}), and (\ref{eq:Omega}). Because the latter are fully determined by the  metric functions, the relativistic contributions to the disk structure and radiation output are directly controlled by the underlying space-time geometry, within the set of assumptions of the NT model. The time-averaged radiant energy flux $F(r)$ emitted from the disk surface can be expressed as~\cite{novikov1973astrophysics,page1974disk}
\begin{equation}
F(r)=
-\frac{\dot{M}_0}{4\pi\sqrt{-g}}\,
\frac{\Omega'}
{\left(\widetilde{E}-\Omega\widetilde{L}\right)^2}
\int_{r_{\mathrm{ms}}}^{r}
\left(\widetilde{E}-\Omega\widetilde{L}\right)
\widetilde{L}'\,dr .
\end{equation}
Here, the lower integration limit corresponds to the marginally stable orbit, $r_{\rm ms}$, which defines the innermost edge of the accretion disk. In turn, the time-averaged torque per unit circumference, $\mathcal{T}_{\phi}^{\phantom{\phi}r}(r)$, due to the presence of stresses in the disk, which is mechanically responsible for the outward transport of angular momentum throughout the disk, is directly related to the radiating flux and given by
\begin{equation}
\mathcal{T}_{\phi}^{\phantom{\phi}r}(r)=-2F(r)
\frac{\widetilde{E}-\Omega\widetilde{L}}{\Omega'}.
\end{equation}
 
Under the assumption of optically thick emission, the luminosity $L(\omega)$ of the disk can be calculated by integrating the blackbody spectral distribution over its surface as~\cite{Torres:2002td}
\begin{equation}
L(\omega)=
\frac{4}{\pi}\,
\cos (i) \, \omega^{3}
\int_{r_{\mathrm{ms}}}^{r_{\mathrm{out}}}
\frac{r\,dr}
{e^{\frac{\omega}{T(r)}}-1},
\end{equation}
where $i$ denotes the inclination angle of the disk relative to the observer, $r_{\mathrm{out}}$ corresponds to the outer radius of the disk, and $\omega$ relates to the radiation frequency $\nu$, via $2\pi\nu=\omega$. To compute this quantity, we rely on the relation between the disk temperature $T(r)$ with the radiant energy flux, $F(r)=\sigma T^4(r)$, where $\sigma$ is the Stefan-Boltzmann constant. Additionally, in our calculations we shall assume the disk to have a face-on orientation (i.e., $i=0$).

\begin{figure*}[t!]
\centering
{
\includegraphics[width=8.07cm]{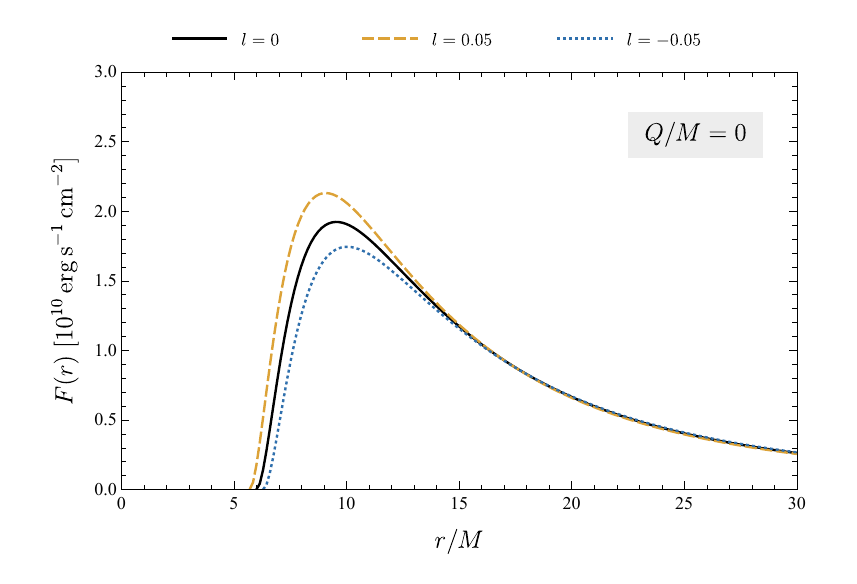}
}
\hspace{0.1cm}
{
\includegraphics[width=8.07cm]{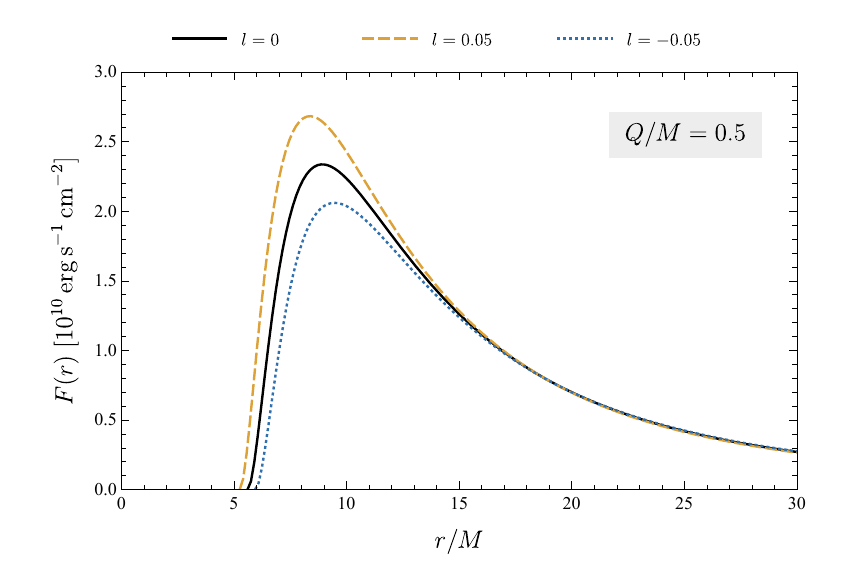}
}
\caption{Time averaged flux $F(r)$ radiated by a thin accretion disk for a RN-like KR black hole of mass $M=4.3 \times 10^6 M_{\astrosun}$ and accretion rate $\dot{M}_0=10^{-8} M_{\astrosun} \, yr^{-1}$. The neutral case $(Q/M=0)$ is depicted in the left plot and the charged black hole $(Q/M=0.5)$ is depicted in the right plot. The solid black curve represents the Schwarzschild and the RN black holes in the left and right plots, respectively. The dashed and dotted curves denote different values of the symmetry-breaking parameter $l$.}\label{fig:ThinAccretionFlux}
\end{figure*} 

Another relevant quantity characterizing the accretion process is the matter-to-radiation conversion efficiency, which measures how the space-time geometry affects the conversion of accretion energy into radiation. This quantity is defined as the ratio between the time-averaged rate at which photon energy is emitted from the disk surface and the rate at which mass-energy is delivered to the central black hole, both measured at infinity~\cite{page1974disk}. Under the assumptions of the NT model and neglecting photon capture by the black hole, this quantity can be expressed in terms of the specific energy of the accreting matter as\footnote{Note that our space-time does not satisfy $g_{\mu\nu} \rightarrow \eta_{\mu\nu}$ at asymptotic infinity. Consequently, the specific energy at infinity reads $\tilde{E}_{\infty}\rightarrow\sqrt{\frac{1}{1-l}}$.}
\begin{equation}
\epsilon=
1-\frac{\widetilde{E}_{\rm ms}}{\widetilde{E}_{\infty}}=
1-\widetilde{E}_{\rm ms}\sqrt{1-l},
\end{equation}
where $\widetilde{E}_{\rm ms}$ corresponds to the specific energy evaluated at the marginally stable orbit and $\widetilde{E}_{\infty}$ corresponds to the asymptotic specific energy.

\begin{figure*}[ht!]
\centering
{\label{fig:Temp_Q0}
\includegraphics[width=8.07cm]{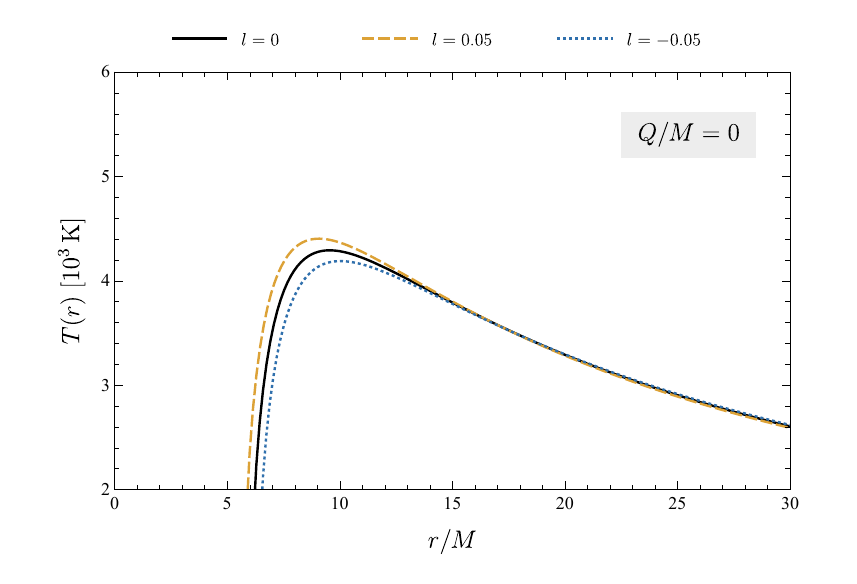}
}
\hspace{0.1cm}
{
\label{fig:Temp_Q0.5}\includegraphics[width=8.07cm]{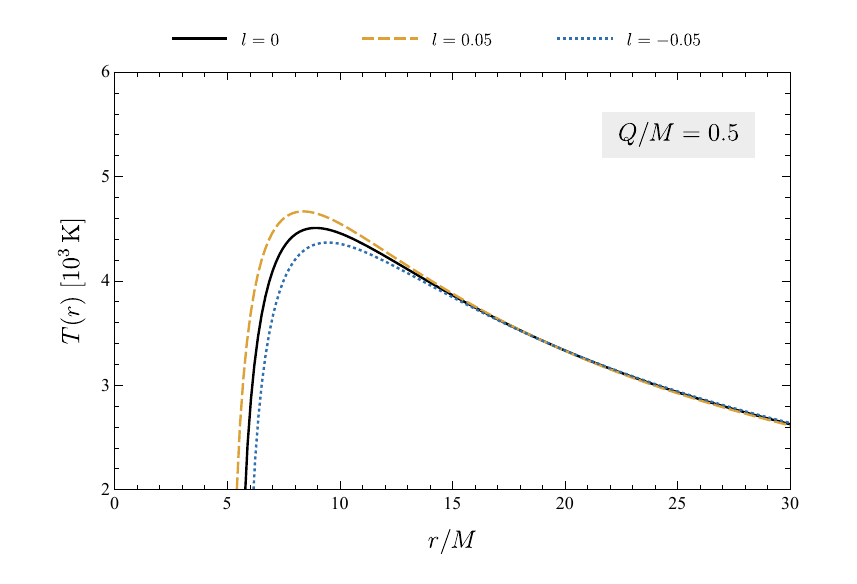}
}
\caption{Disk temperature profiles of the thin accretion disk for a RN-like KR black hole of mass $M=4.3 \times 10^6 M_{\astrosun}$ and accretion rate $\dot{M}_0=10^{-8} M_{\astrosun} \, yr^{-1}$. The neutral case $(Q/M=0)$ is depicted in the left plot and the charged black hole $(Q/M=0.5)$ is depicted in the right plot. The solid black curve represents the Schwarzschild and the RN black holes in the left and right plots, respectively. The dashed and dotted curves denote different values of the symmetry-breaking parameter $l$.}\label{fig:ThinAccretionTemp}
\end{figure*} 

\begin{figure*}[ht!]
\centering
{\label{fig:Spectra_Q0}
\includegraphics[width=8.07cm]{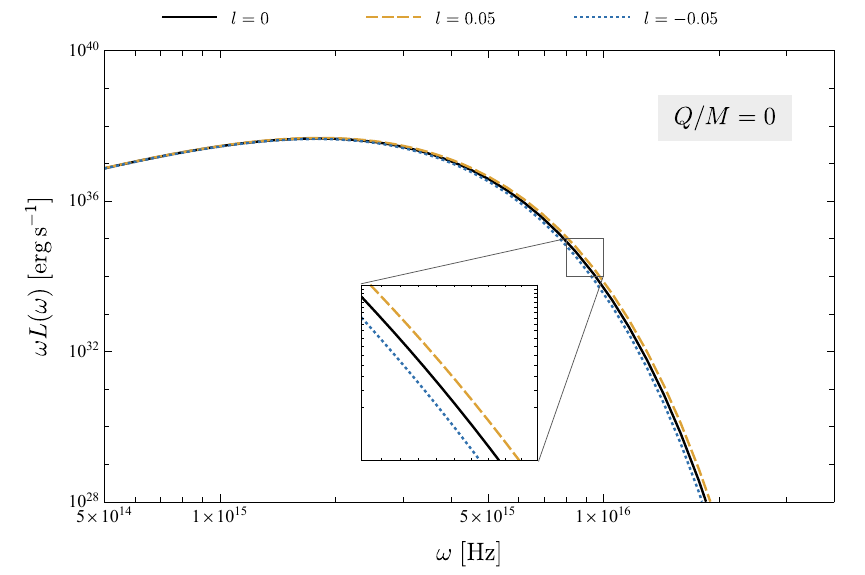}
}
\hspace{0.1cm}
{
\label{fig:Spectra_Q0.5}\includegraphics[width=8.07cm]{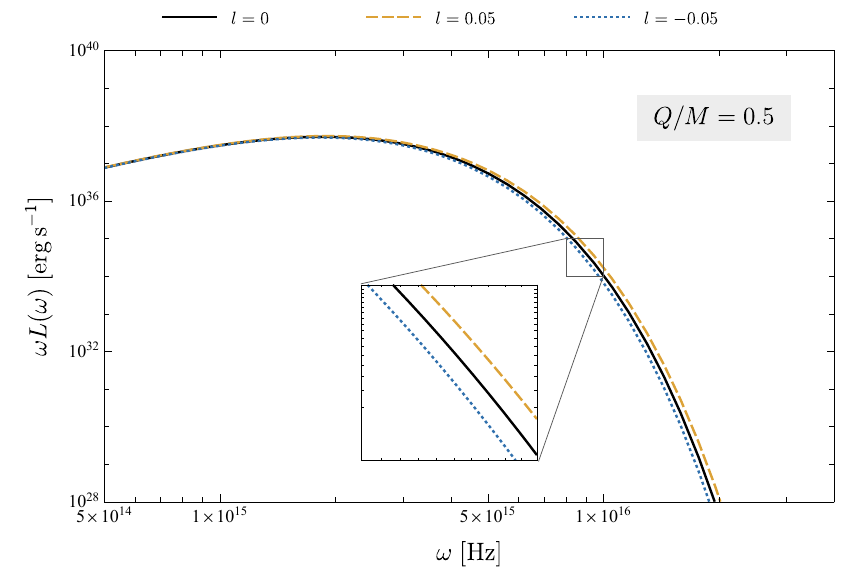}
}
\caption{Emission spectra $\omega L(\omega)$ of a thin accretion disk surrounding a RN-like KR black hole of mass $M=4.3 \times 10^6 M_{\astrosun}$ and accretion rate $\dot{M}_0=10^{-8} M_{\astrosun} \, yr^{-1}$. The neutral case $(Q/M=0)$ is depicted in the left plot and the charged black hole $(Q/M=0.5)$ is depicted in the right plot. The solid black curve represents the Schwarzschild and the RN black holes in the left and right plots,respectively. The dashed and dotted curves denote different values of the symmetry-breaking parameter $l$.}\label{fig:ThinAccretionSpectra}
\end{figure*} 

\begin{figure*}[ht!]
\centering
{\label{fig:Torque_Q0}
\includegraphics[width=8.07cm]{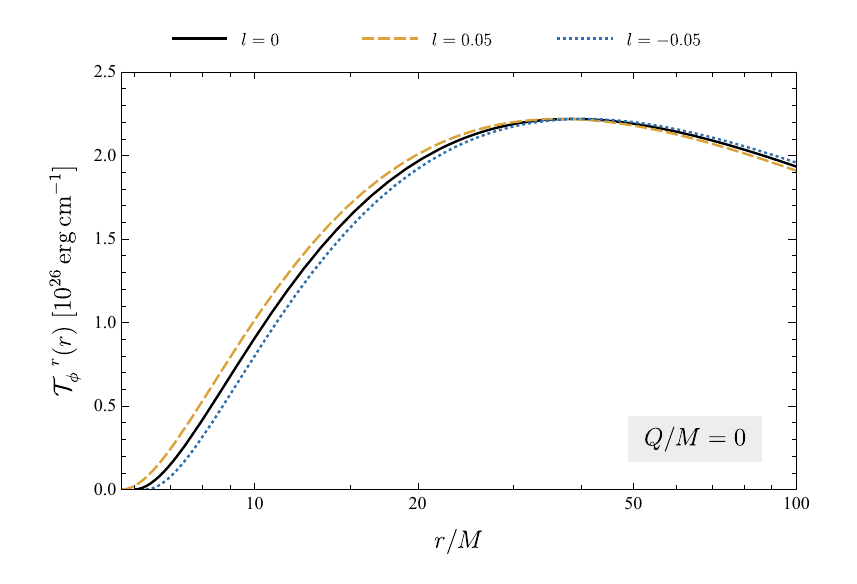}
}
\hspace{0.1cm}
{
\label{fig:Torque_Q0.5}\includegraphics[width=8.07cm]{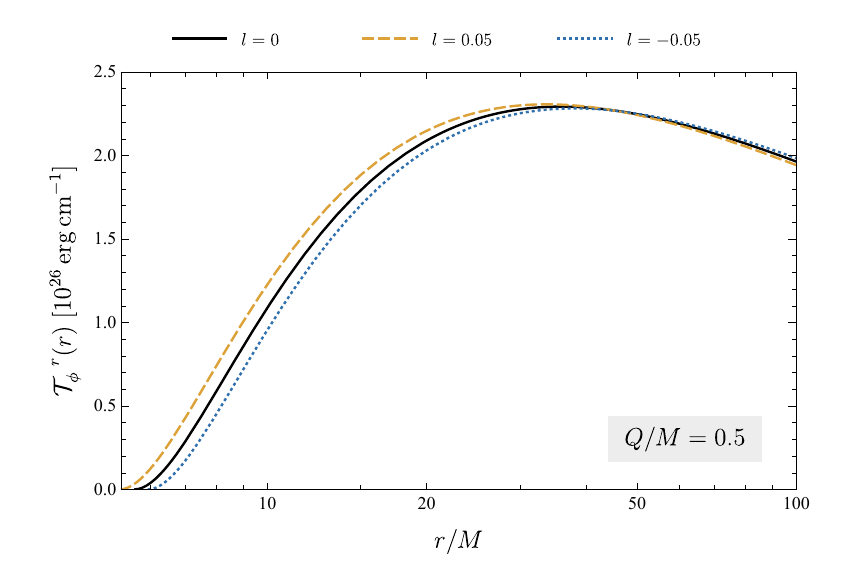}
}
\caption{Time-averaged torque $\mathcal{T}_{\phi}^{\phantom{\phi}r}(r)$ radiated by a thin disk for a RN-like KR black hole of mass $M=4.3 \times 10^6 M_{\astrosun}$  and accretion rate $\dot{M}_0=10^{-8} M_{\astrosun} \, yr^{-1}$. The neutral case $(Q/M=0)$ is depicted in the left plot and the charged black hole $(Q/M=0.5)$ is depicted in right plot. The solid black curve represents the Schwarzschild and the RN black holes in the left and right plots, respectively. The dashed and dotted curves denote different values of the symmetry-breaking parameter $l$.}\label{fig:ThinAccretionTorque}
\end{figure*} 

In Figs. \ref{fig:ThinAccretionFlux}-\ref{fig:ThinAccretionTorque}, we plot the energy flux, disk temperature, emission spectrum, and torque, respectively, of an accretion disk with a mass accretion rate of $\dot{M}_0=10^{-8} M_{\astrosun} \, \text{yr}^{-1}$ around a charged KR black hole with mass $M=4.3 \times 10^6 M_{\astrosun}$ (the estimated one for Sgr A$^*$). We study both neutral and charged scenarios, $Q/M = \{0 ; 0.5\}$, and three configurations for the symmetry-breaking parameter $l=\{-0.05; 0; 0.05\}$. The radial coordinate for the innermost stable circular orbits (ISCO) $r_{\rm ms}$, corresponding to the innermost edge of the disk, is obtained from the marginal stability condition $\frac{d\widetilde{L}^{\,2}}{dr}=0$ (or, equivalently, from $\frac{d^2 V_{eff}}{dr^2}=0$)
which, defining $k=1-l$, yields
\begin{eqnarray}
    - k^3 M r^2 (r-6 k M)-9 k^2 M Q^2 r +4 Q^4 =0.
\end{eqnarray}
Within the non-extremal black hole parameter domain considered here, the previous equation possesses a unique physically relevant outer root, which defines the radius $r_{\rm ms}$ corresponding to the innermost edge of the accretion disk.

Comparing the radiant energy flux profiles, depicted in Fig.~\ref{fig:ThinAccretionFlux}, we see that the intensity of the flux emerging from the disk surface increases with the charge. This shifts the inner edge of the disk towards smaller radii, allowing the accretion flow to release gravitational energy in a region closer to the black hole and resulting in larger flux maxima. In both the neutral and charged scenarios, positive values of $l$ lead to an increase in the emitted flux, while negative values suppress the radiative output, an effect that is further amplified in the charged space-time. This pattern also manifests in the radial temperature profiles in Fig.~\ref{fig:ThinAccretionTemp}.

In the disk's intrinsic emission spectra $\omega L(\omega)$, plotted in Fig.~\ref{fig:ThinAccretionSpectra}, the differences between the profiles remain small around the spectral maximum and become more pronounced along the high-frequency tail, in the ultraviolet regime and beyond. As can be seen in zoom-in boxes in each plot, the spectrum extends towards higher frequencies for positive values of $l$. In contrast, negative values of $l$ cause the emission spectra to fall-off earlier. This effect is enhanced in the charged configuration, reflecting the larger differences in the temperature of the inner disk.

The time-averaged torque per unit circumference profiles are displayed in Fig.~\ref{fig:ThinAccretionTorque}. In both plots, the angular momentum transport progressively increases, reaching a maximum at an intermediate radius. Beyond this point, the decreasing shear in the outer regions of the disk reduces the local stresses, although the outward transport of angular momentum required for accretion is maintained. The LIV effects shift the radial profile of the torque. Positive values of $l$ result in both an earlier increase and decline in the torque profile, whereas negative values delay the profile trend. This behavior is clearer in the charged case, where positive values of $l$ result in a slightly larger maximum, while negative values reduce the peak and shift it towards larger radii. As with the previous disk parameters, the inclusion of electric charge makes the torque profile more sensitive to LIV effects, amplifying the dependence of both the position and magnitude of the peak on $l$.

\begin{figure}[ht!]
    \centering
    \includegraphics[width=\columnwidth]{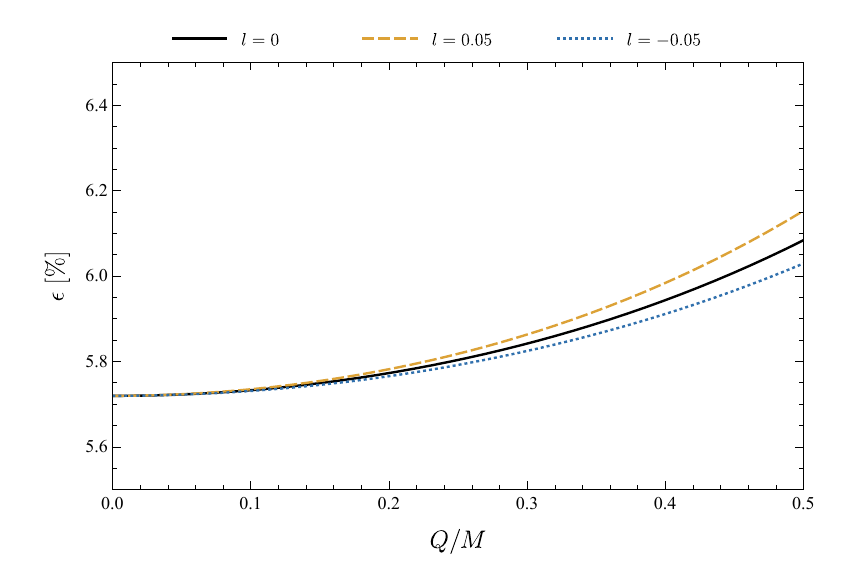}
    \caption{Conversion efficiency $\epsilon$, in percentage, as a function of the charge, for a charged KR black hole~\eqref{eq:line_element}--\eqref{eq:line_element3}. The solid black curve represents the Reissner--Nordstr\"om black hole.}
    \label{fig:Efficiency}
\end{figure}

The radiative conversion efficiency of the accretion disk is shown in Fig.~\ref{fig:Efficiency}. We observe that the efficiency increases with the charge parameter for all values of $l$, reflecting the inward shift of the marginally stable orbit $r_{\rm ms}$. The Lorentz symmetry-breaking parameter also modifies the efficiency, with positive values of $l$ resulting in higher conversion efficiencies, while negative values suppress the radiative output. This behavior is consistent with the previous results for the disk properties, showing that the joint effects of $Q$ and $l$ become increasingly relevant in the strong-field region. 

\section{Thick Accretion Disks} \label{IV}

We now consider a relativistic thick accretion disk modeled as a perfect fluid in stationary and axisymmetric equilibrium. The fluid undergoes purely circular motion, with four-velocity $u^\mu=(u^t, 0, 0, u^\phi)$ satisfying $g_{\mu\nu}u^\mu u^\nu=-1$. Assuming a barotropic equation of state $p=p(e)$, where $p$ is  the pressure and $e$ the energy density, we define the rest mass density $\rho$, the enthalpy density $\omega=e+p$, and the specific enthalpy $h=\omega/\rho$. 

For thick accretion disks, we define the specific angular momentum per unit energy of purely circular fluid motion as
\begin{equation}\label{eq:fluid_ell}
\ell(r,\theta)\equiv
-\frac{u_{\phi}}{u_t}
=
\frac{C(r)\sin^2\theta}{A(r)}\Omega,
\end{equation}
with $\Omega=u^{\phi}/u^t$. Using the condition
$V_{{\rm eff}}'=0$ allows one to obtain the effective angular momentum of a test particle moving along circular geodesics
\begin{equation}\label{eq:eff_L}
\ell_{\rm eff}(r)
=
\frac{\widetilde L}{\widetilde E}
=
\frac{C(r)}{A(r)}\sqrt{\frac{A'(r)}{C'(r)}},
\end{equation}

The characteristic radii governing thick-disk configurations are the marginally stable orbit $(r_{\rm ms})$ and the marginally bound orbit $(r_{\rm mb})$. Since the Killing time used in Eq.~\eqref{eq:generic_metric} is not normalized to unit lapse at infinity, one has $\widetilde E_{\infty}=1/\sqrt{1-l}$. Accordingly, the marginally bound orbit is determined by
\begin{equation}
\widetilde E(r_{\rm mb})=\widetilde E_{\infty}=\frac{1}{\sqrt{1-l}},
\end{equation}
or, equivalently, by $\bar E(r_{\rm mb})=1$ for the normalized energy $\bar E\equiv\sqrt{1-l}\,\widetilde E$. The associated angular momenta are then given by $\ell_{\rm ms}=(\widetilde L/\widetilde E)_{r_{\rm ms}}$ and $\ell_{\rm mb}=(\widetilde L/\widetilde E)_{r_{\rm mb}}$.

\begin{figure*}[t!]
\centering
\subfigure[] 
{\label{fig:ell_eff}\includegraphics[width=8.5cm]{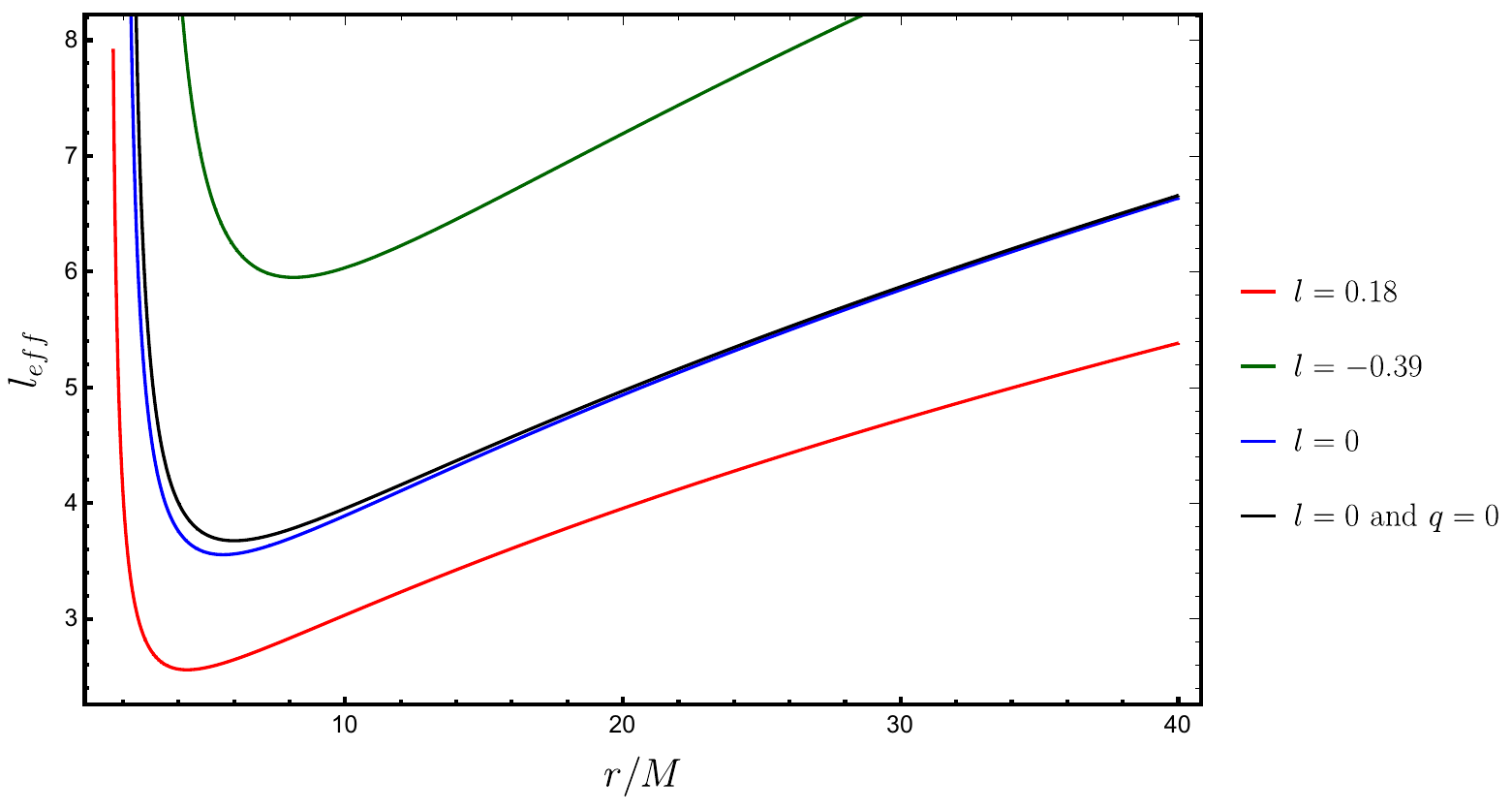} }
\hspace{0.1cm}
\subfigure[] 
{\label{fig:ISCO_hor}\includegraphics[width=8.cm]{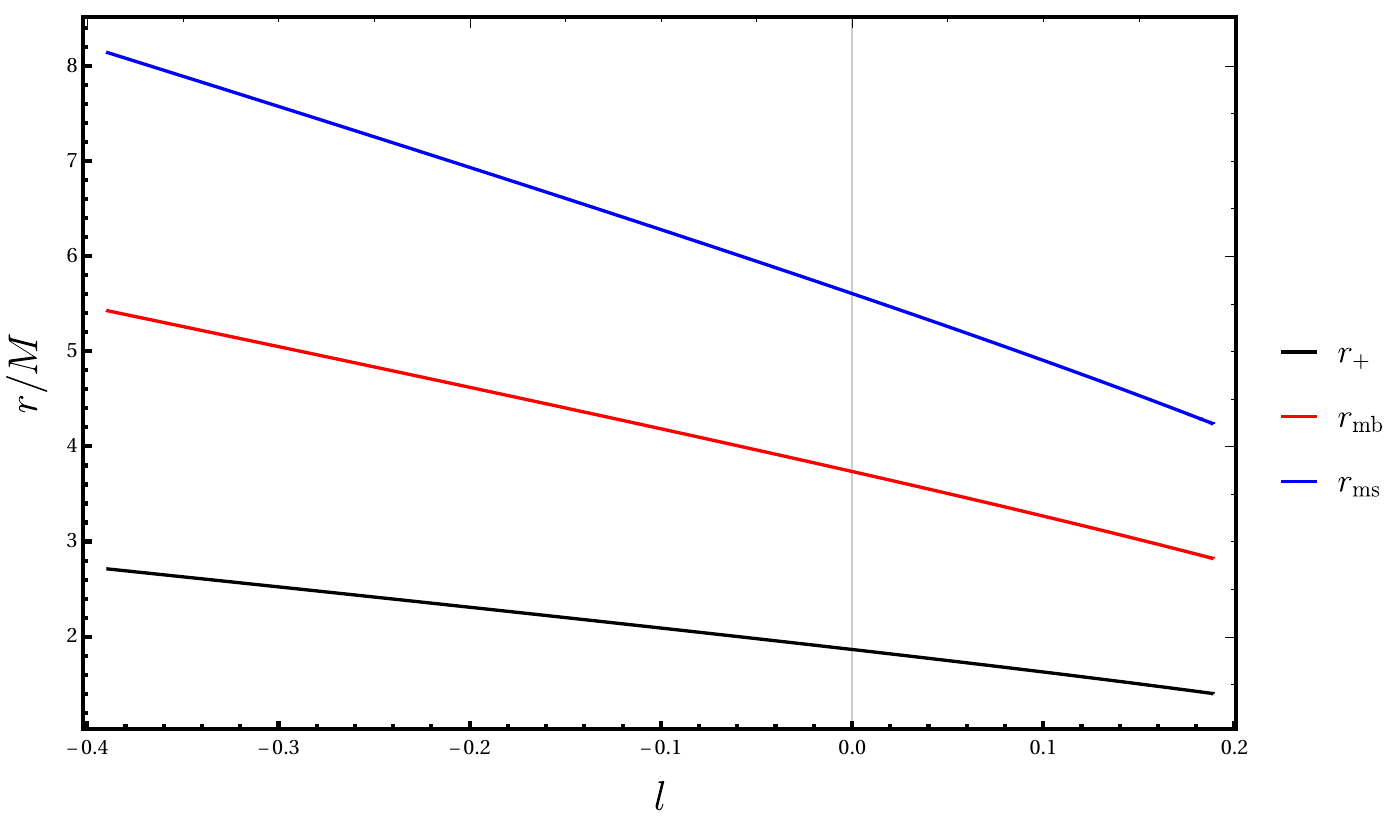} }
\caption{(a): Effective angular momentum $\ell_{\rm eff}(r)$ for different values of the Lorentz symmetry-breaking parameter $l$. (b): Radial dependence of the characteristic radii $r_+$, $r_{\rm mb}$, and $r_{\rm ms}$ as a function of $l$.}
\end{figure*}

In Fig~\ref{fig:ell_eff} we assess how the effective angular momentum behaves with respect to the symmetry-breaking parameter $l$ of a charged KR black hole, when plotted as a function of the radial coordinate. The curves start at the position of the outer horizon $r_+$, which depends explicitly on $l$, indicating that this parameter modifies the horizon location and therefore the accessible region for circular motion. For negative values of $l$, the horizon is shifted to larger radii and the effective angular momentum required for circular motion increases near the black hole, indicating a deeper effective gravitational potential. In contrast, positive values of $l$ shift the horizon inward and reduce the required angular momentum close to the horizon, suggesting a weakening of the effective gravitational potential. At large distances, all curves tend to converge, recovering the expected asymptotic behavior.

Figure~\ref{fig:ISCO_hor} shows how the orbital structure of the black hole is modified by the $l$ parameter. Within the allowed interval of $l$, positive values $l>0$ reduce the outer horizon $r_{+}$. Likewise, both $r_{\rm mb}$ and $r_{\rm ms}$ are shifted toward smaller radii, allowing such orbits to exist closer to the black hole. For $l<0$, the outer horizon increases and both $r_{\rm mb}$ and $r_{\rm ms}$ are shifted toward larger radii, pushing these orbits farther away from the black hole. The usual ordering $r_+<r_{\rm mb}<r_{\rm ms}$ is preserved throughout the parameter range, ensuring the physical consistency of the solution.  In the limit $Q/M=0$ and $l=0$, the standard Schwarzschild results $r_+=2M$, $r_{\rm mb}=4M$, and $r_{\rm ms}=6M$ are recovered.

In the case of a barotropic equation of state, the equi-$\ell$ and equi-$\Omega$ surfaces coincide, defining the so-called von Zeipel cylinders \cite{Abramowicz1974}.  Thus, once the metric and the relation between $\ell$ and $\Omega$ are specified, the von Zeipel surfaces are fixed. To this end, we consider a perfect fluid whose energy–momentum tensor is given by
\begin{eqnarray}
    T^{\mu\nu}=(e+p)u^\mu u^\nu+p g^{\mu\nu}\,.
\end{eqnarray}
where $e$ is the total energy density of the fluid and $p$ its pressure, consistently with the notation introduced above; $\rho$ denotes the rest-mass density. From the conservation law $\nabla_\mu T^{\mu\nu}=0$, the  equations describing the structure of a perfect fluid are given by \cite{ Abramowicz1974,  PhysRevD.110.124030, Font:2002bi,  PaczynskiAbramowicz1982, Chen:2024jsv},
\begin{eqnarray}
  -\nabla_\mu \ln(-u_t)+\frac{\Omega \nabla_\mu \ell}{1-\ell\Omega}=\frac{\nabla_\mu p}{e+p}
\end{eqnarray}
where we have made use of Eqs.\,\eqref{eq:Conserved_EandL}. In accordance with the von Zeipel theorem \cite{Abramowicz1974}, the relation $\Omega=\Omega(\ell)$ holds, such that we can obtain
\begin{eqnarray}
W-W_{\rm in}=\ln(-u_t)-\ln(-u_t)_{\rm in}-\int^\ell_{\ell_{\rm in}}\left(\frac{\Omega}{1-\Omega \ell' }\right) d\ell' \label{W1}
\end{eqnarray}
where the subscript ``in" denotes evaluation at the inner edge of the disk.  Thus, in the case of a barotropic fluid, the surfaces of constant pressure (isobars) coincide with the equipotential surfaces $W(r,\theta)={\rm const}$, as prescribed by Boyer’s condition \cite{Boyer1965}. These equipotential surfaces $W(r, \theta)$ are then readily obtained from Eq.\,\eqref{W1}, taking into account that $u_t$ can be expressed in terms of the metric functions and the angular velocity. In the simplest case of a torus with a uniform effective angular momentum distribution \begin{eqnarray}
\ell(r, \theta)=\ell_0,
\end{eqnarray}
the potential $W(r,\theta)$ takes the form given by \cite{PhysRevD.110.124030}
\begin{eqnarray}
    W(r, \theta)=\frac{1}{2}\ln\left(\frac{A(r)C(r)\sin^2\theta}{C(r)\sin^2\theta-A(r)\ell^2_0}\right)\,.  \label{eq:W2}
\end{eqnarray}
The potential \eqref{eq:W2} fully determines the equilibrium structure of the torus, with its critical points and level surfaces identifying the disk center, cusp, and outer boundary.

\subsection{Constant angular momentum tori around a charged KR black hole}

We now apply the above framework to the charged KR solution given by Eqs. (\ref{eq:generic_metric})-(\ref{eq:line_element3}). Neglecting the disk's self-gravity, and adopting a constant effective angular momentum $\ell_0$, the potential in Eq.~\eqref{eq:W2} determines the family of equipotential surfaces. A specific fluid torus is selected only after choosing the surface potential (or, equivalently, a filling parameter), so a fixed $\ell_0$ by itself does not uniquely determine the outer boundary of the fluid. The potential defines open or closed equipotential surfaces in the $(r, \theta)$ plane, with the closed ones providing stationary fluid configurations when an appropriate filling level is chosen and the fluid region does not enclose the event horizon. The pressure vanishes at the chosen torus surface, while its gradient remains nonzero in general.

As shown in Fig.~\ref{fig:ell_eff}, both $r_{\rm mb}$ and the effective angular momentum depend explicitly on the parameter $l$ and on the $Q/M$ ratio. This dependence produces quantitative differences in the equipotential surfaces with respect to the Schwarzschild and RN cases \cite{Font:2002bi, Chen:2024jsv, PhysRevD.110.124030}. Hence, for each value of the parameter $l$, the potential $W(r,\theta)$ is determined by a specific value of $\ell_0$ lying between $\ell_{\rm ms}$ and $\ell_{\rm mb}$, giving rise to closed equipotential surfaces with a cusp and a center, and consequently to quantitatively distinct thick-disk configurations.

\begin{figure}[ht!]
\centering
\subfigure[] 
{\label{fig2a}\includegraphics[width=8.cm]{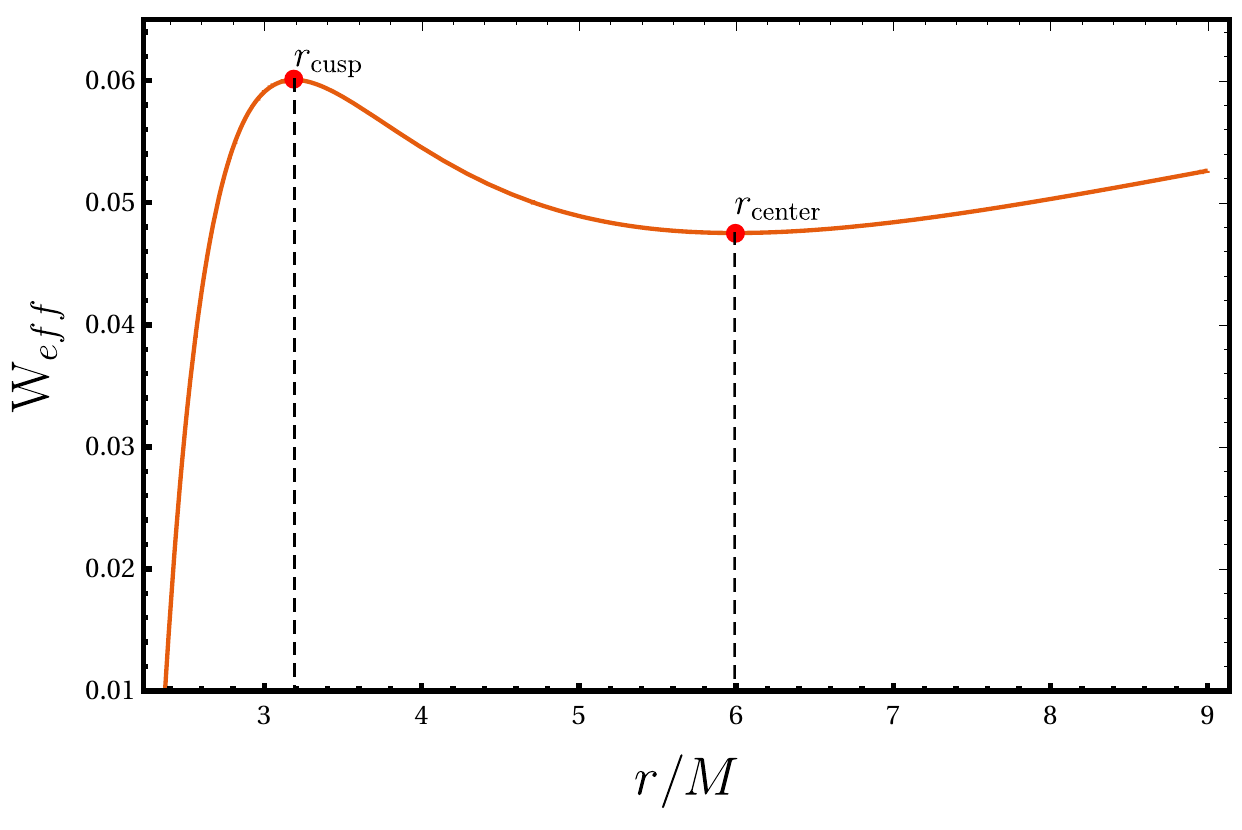} }
\hspace{0.1cm}
\subfigure[] 
{\label{fig2b}\includegraphics[width=8.cm]{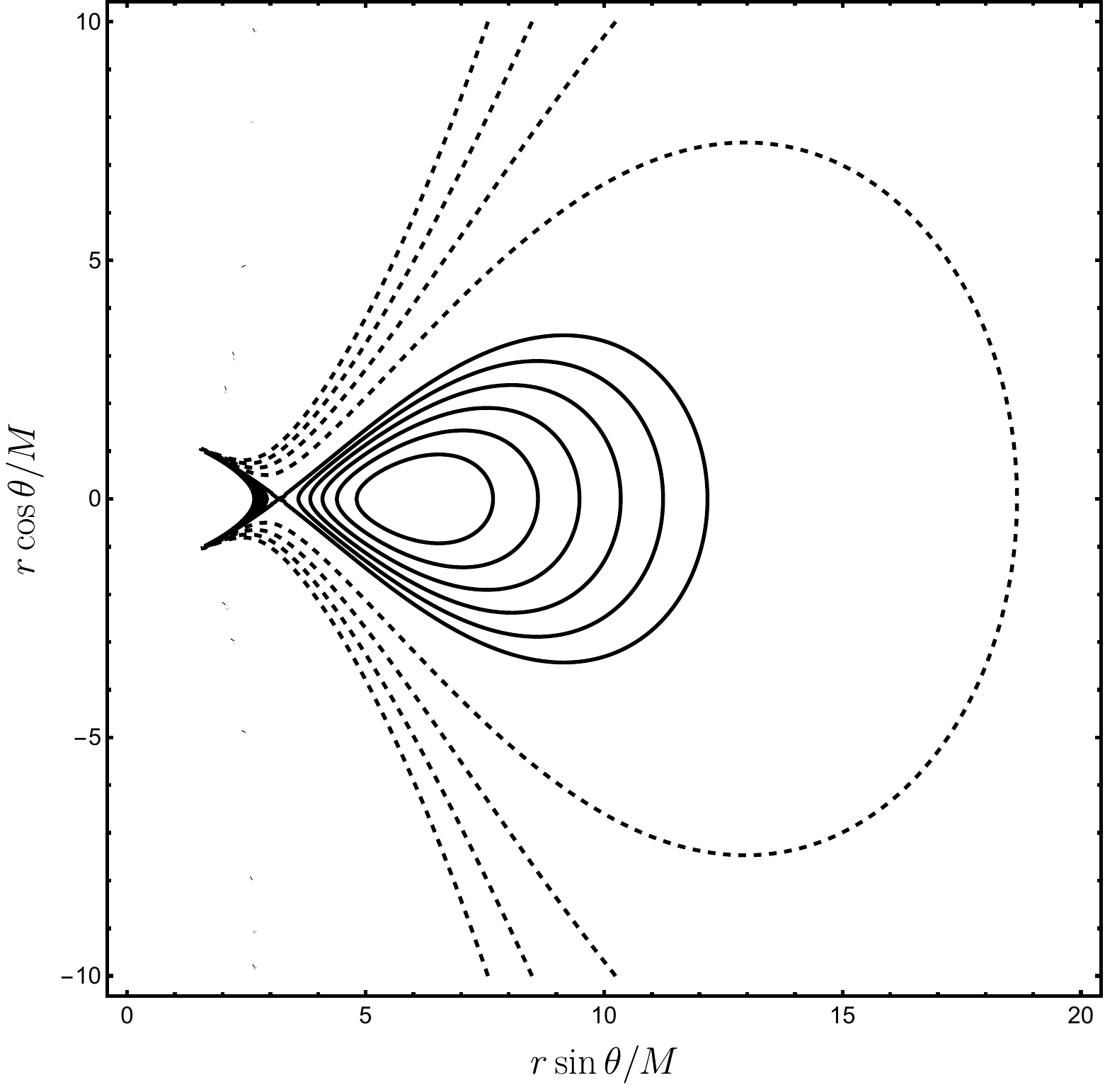} }
\caption{(a): Effective potential \eqref{eq:W2} for $l=0.189$, $Q/M=0.5$, and $\ell_0=2.6$, with the cusp and center of the torus identified. (b): Corresponding fluid topology.}\label{Fig2}
\end{figure}

In order to perform a meaningful assessment of the LIV effects, we can define $\ell_0$ such that it occupies the same relative position with respect to $\ell_{\rm ms}$ and $\ell_{\rm mb}$ for the Schwarzschild, RN, and charged KR metrics, namely
\begin{eqnarray}
    \ell_0=\ell_{\rm ms}+\eta(\ell_{\rm mb}-\ell_{\rm ms})\,,
\end{eqnarray}
with $\eta$ taking values in different intervals. For instance, for $l=0.189$ and $Q/M=0.5$, we obtain $\ell_{\rm ms}=2.50657$ and $\ell_{\rm mb}=2.72719$. As a first case, we retain the value $\ell_0=2.6$ used in the numerical configuration. With the quoted values of $\ell_{\rm ms}$ and $\ell_{\rm mb}$, this corresponds to $\eta\simeq0.4235$, and therefore lies in the interval $\ell_{\rm ms}<\ell_0<\ell_{\rm mb}$. Plugging this value into Eq.~\eqref{eq:W2}, Fig.~\ref{fig2a} shows a cusp located at $r_{\rm cusp}=3.18$ and a stable centre at $r_{\rm center}=5.99$, indicating the hydrostatic equilibrium of the fluid. As shown by the solid curves in Fig.~\ref{fig2b}, there is a closed torus with a well-defined potential, demonstrating that the fluid is confined to a finite spatial region and cannot escape radially either toward infinity or directly into the black hole, except through the cusp. The pressure vanishes at the outer boundary of the torus, sustaining the disk without the need for viscosity. The dashed curve denotes $W(\infty)= 0.066$, corresponding to the asymptotic limit of the effective potential and determining the maximum outer extension of the torus. For $W>0.066$ equipotential surfaces become open and matter is no longer gravitationally bound, leading to outflow toward infinity.  

\begin{figure}[ht!]
\centering
\subfigure[] 
{\label{fig3a}\includegraphics[width=8.cm]{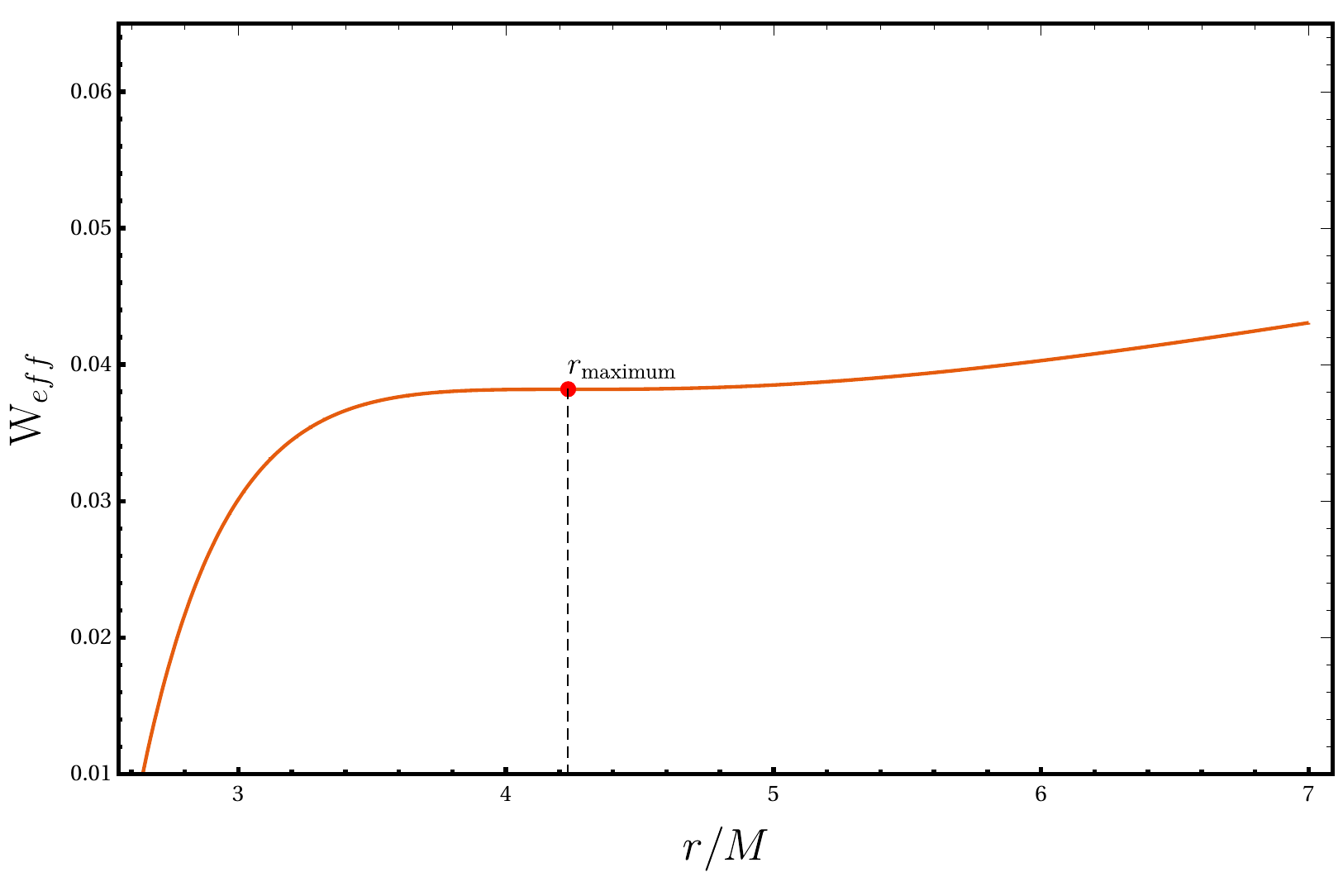} }
\hspace{0.1cm}
\subfigure[] 
{\label{fig3b}\includegraphics[width=8.cm]{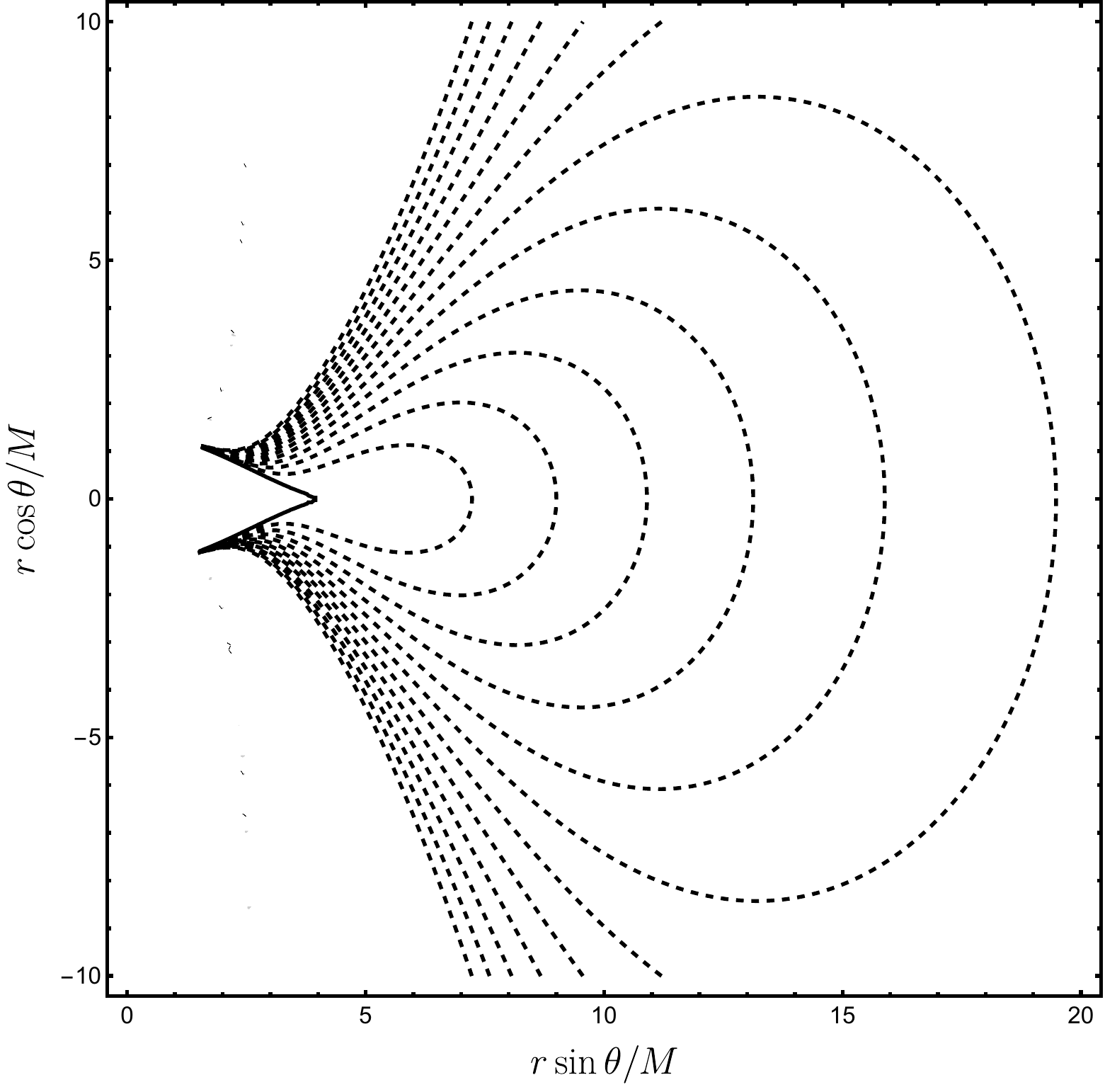} }
\caption{(a): Effective potential \eqref{eq:W2} for  $l=0.189$ and $Q/M=0.5$, with $\ell_0=\ell_{\rm ms}$ and $r_{\rm maximum}$ identified. (b): Corresponding fluid topology.}\label{Fig3}
\end{figure}

For the marginally stable configuration, $\ell_0 = \ell_{\rm ms}$, the  potential displays only a single local maximum at $r=4.23122$ and no minimum, as can be seen in Fig.~\ref{fig3a}. The absence of a degenerate center indicates that the confining region of the torus has already collapsed at $\ell_0=\ell_{\rm ms}$. This point corresponds to an unstable radial equilibrium where gravitational and centrifugal effects momentarily balance. However, since no local minimum is present, there is no confining potential well capable of sustaining a stationary toroidal configuration. Therefore, this maximum does not represent a functional cusp associated with a finite disk, but rather an isolated unstable critical point. Its presence indicates that the marginally stable configuration has already lost the ability to support hydrostatic equilibrium. The LIV correction does not introduce a new marginal topology relative to the RN case. In this restricted stationary analysis, the $l$ parameter shifts the location and scale of the marginal configuration. The disappearance of a confining potential well characterizes a change in equilibrium topology, but it should not by itself be interpreted as proof of a time-dependent runaway instability. The equipotential curves for this case are shown in Fig.~\ref{fig3b}; the dashed curves indicate an unstable fluid.

\begin{figure}[t!]
\centering
\subfigure[] 
{\label{fig4a}\includegraphics[width=8.cm]{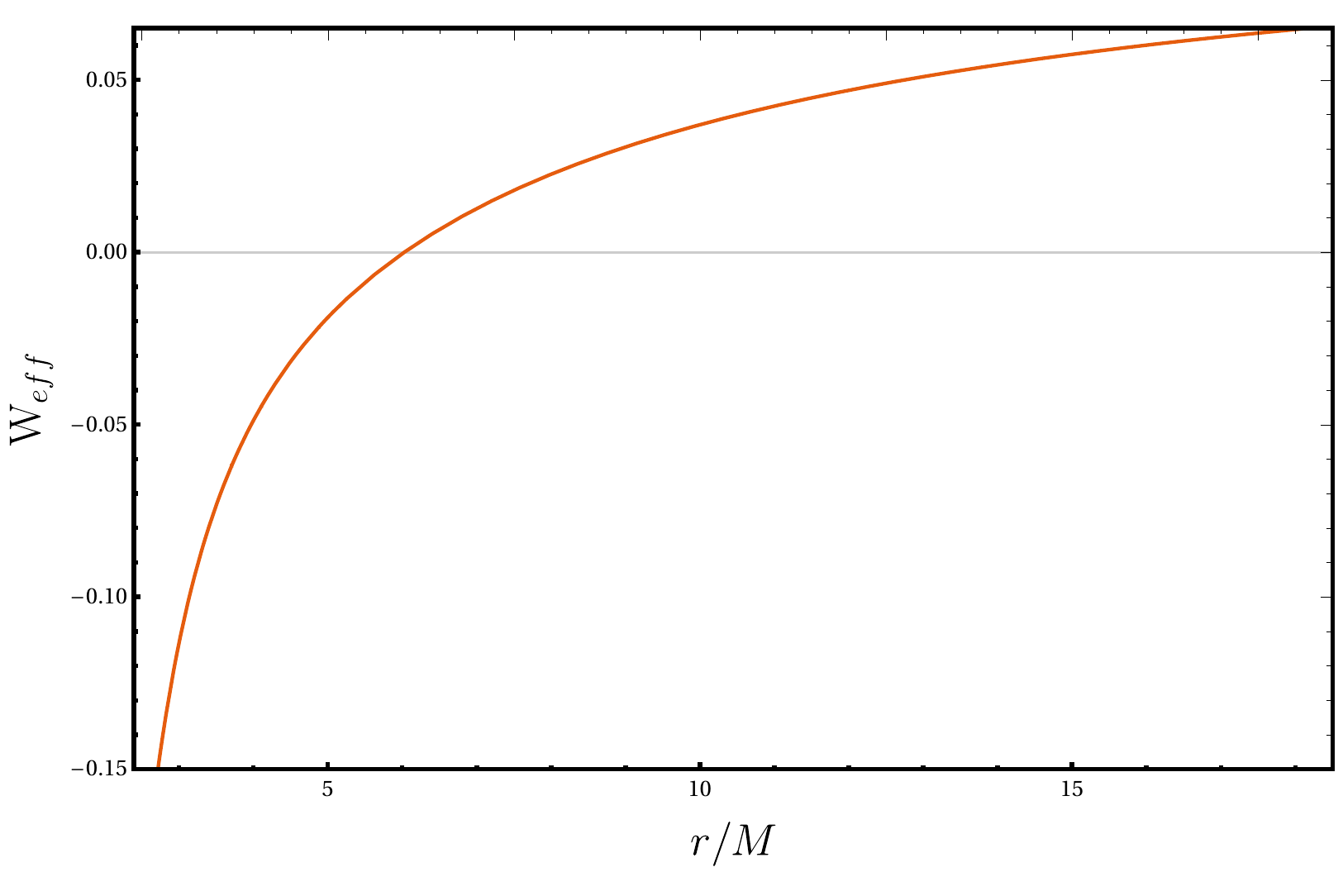} }
\hspace{0.1cm}
\subfigure[] 
{\label{fig4b}\includegraphics[width=8.cm]{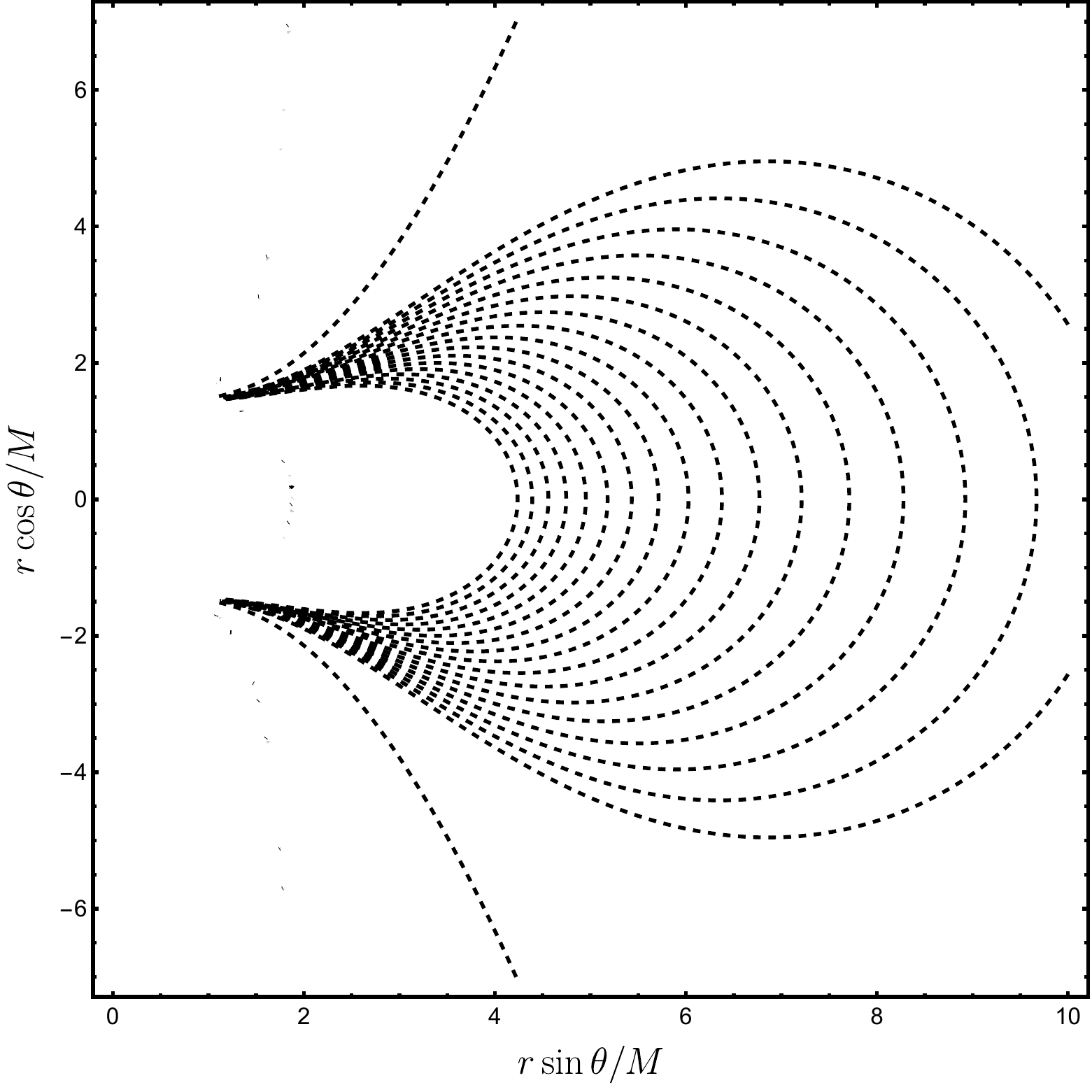} }
\caption{(a): Effective potential \eqref{eq:W2} for  $l=0.189$ and $Q/M=0.5$, with $\ell_0<\ell_{\rm ms}$. (b): Corresponding fluid topology.}\label{Fig4}
\end{figure}

In the $\ell_0<\ell_{\rm ms}$ scenario, the critical point is shifted inside $r_{\rm ms}$; therefore, no cusp or center is present, and the critical equipotential configuration displayed in Fig.~\ref{Fig4} is open and unconfined. In this regime, the plotted equipotential surfaces do not define a finite cusp-bearing torus. As can be seen in Table~\ref{TAB1}, all three geometries exhibit open configurations for $\ell_0 < \ell_{\rm ms}$. The progressive inward shift of the unstable critical point and the steepening of the potential in the charged KR geometry show that LIV effects modify the stationary equilibrium topology more strongly than in the Schwarzschild and RN cases. A statement about dynamical growth rates would require a time-dependent hydrodynamical analysis.

\begin{figure}[t!]
\centering
\subfigure[] 
{\label{fig5a}\includegraphics[width=8.cm]{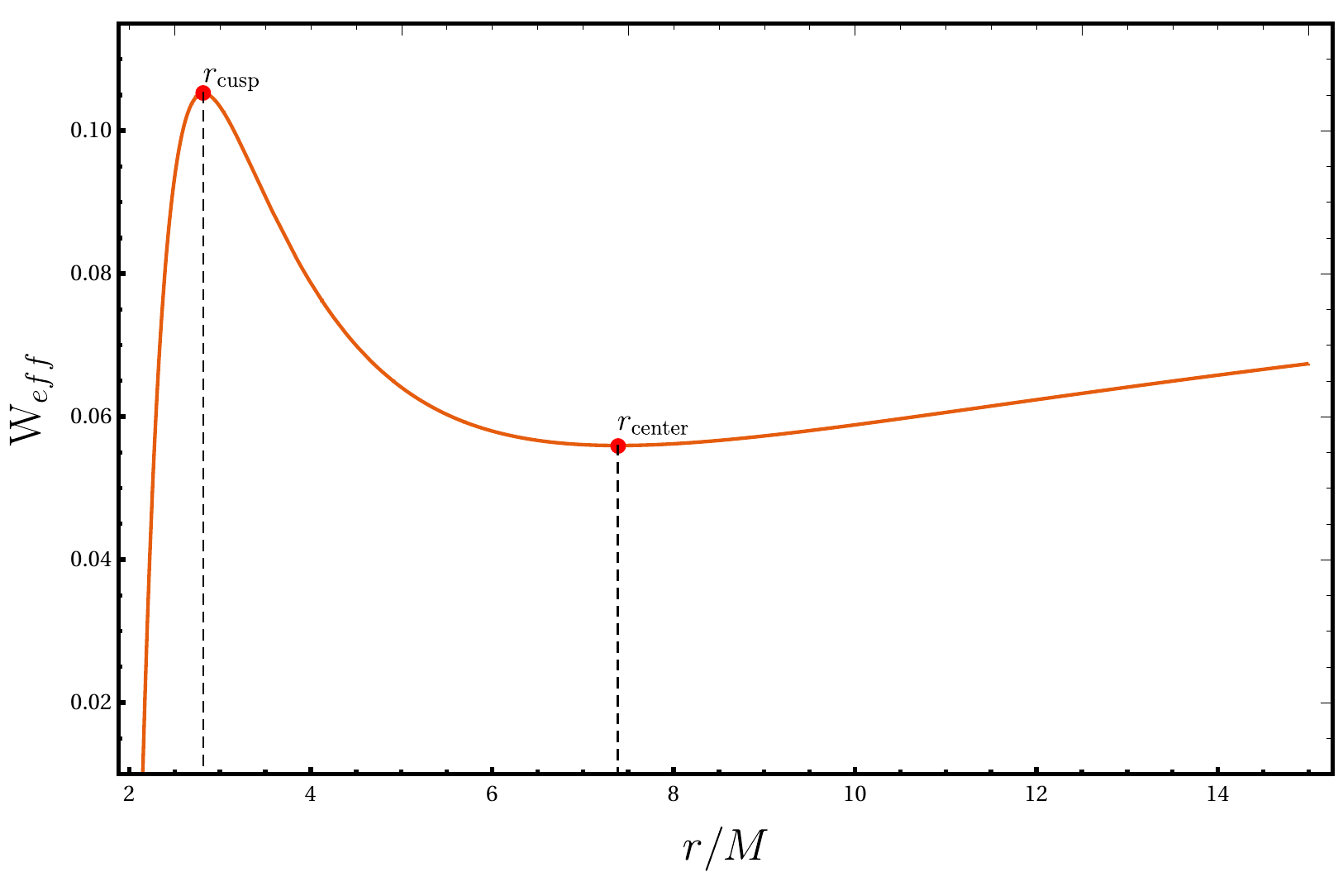} }
\hspace{0.1cm}
\subfigure[] 
{\label{fig5b}\includegraphics[width=8.cm]{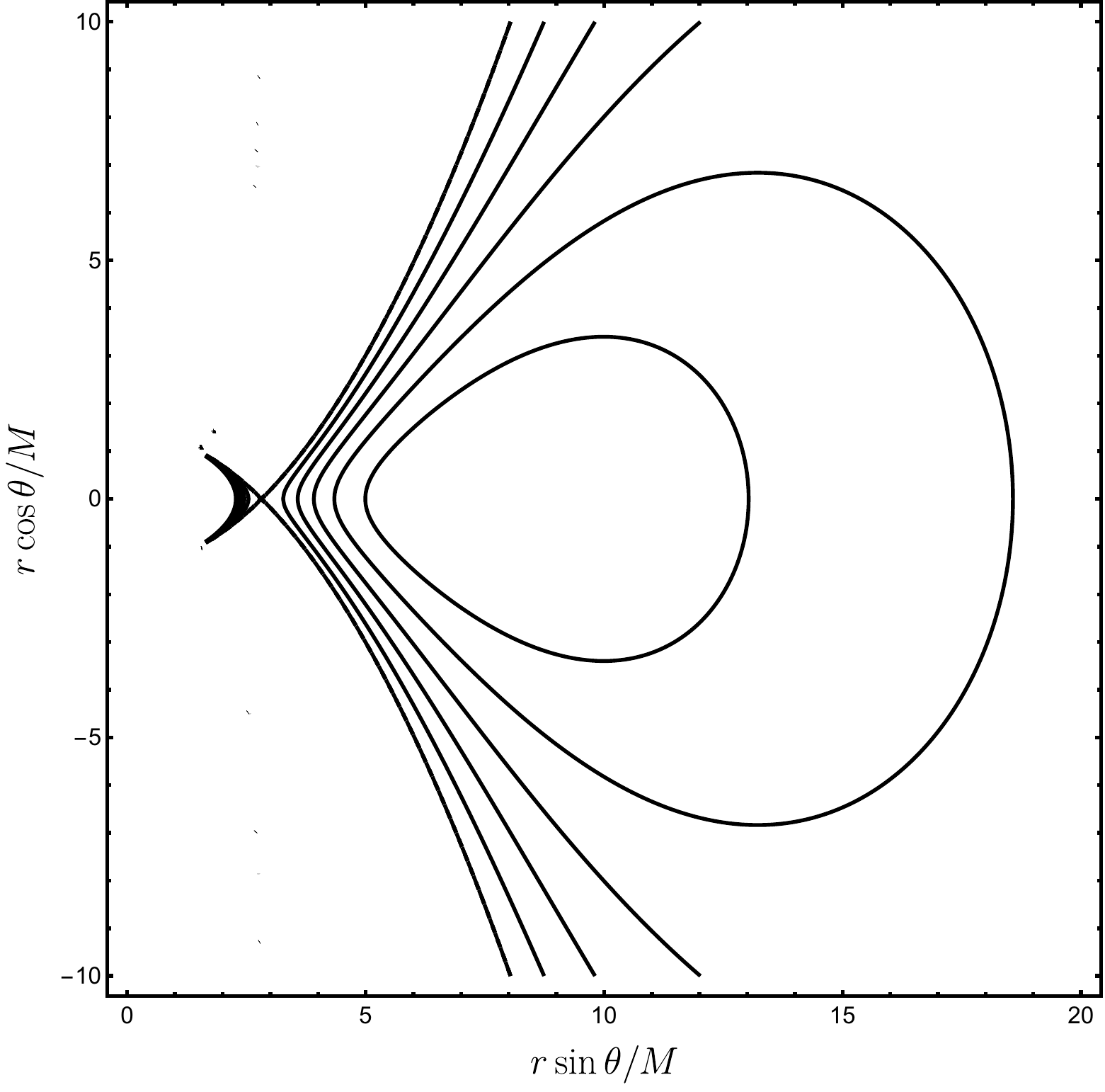} }
\caption{(a): Effective potential \eqref{eq:W2} for  $l=0.189$ and $Q/M=0.5$, with $\ell_0=\ell_{\rm mb}$ and the cusp and torus centre identified. (b): Corresponding fluid topology.}\label{Fig5}
\end{figure}

For $\ell_0=\ell_{\rm mb}$, there is a center located at $r_{\rm center}=7.38432$ and a cusp at $r_{\rm cusp}=2.81762$, as can be seen in Fig.~\ref{fig5a},  while the outer equipotential surface reaches the asymptotic level $W(\infty)=0.105228$, as evidenced in Fig.~\ref{fig5b}, implying an infinitely-long disk. In the charged KR space-time, the reduced value of $\ell_{\rm mb}$ shows that marginally bound configurations are attained for smaller angular momentum, reflecting a weakened gravitational confinement. The fact that the outer critical equipotential (dashed curve in Fig.~\ref{fig5b}) passes through the cusp indicates that $W_{\rm cusp}=W(\infty)$ means that the fluid at the cusp is marginally bound and the torus becomes infinitely extended, marking the transition between confined and unbound configurations.

\begin{figure}[t!]
\centering
\subfigure[] 
{\label{fig6a}\includegraphics[width=8.cm]{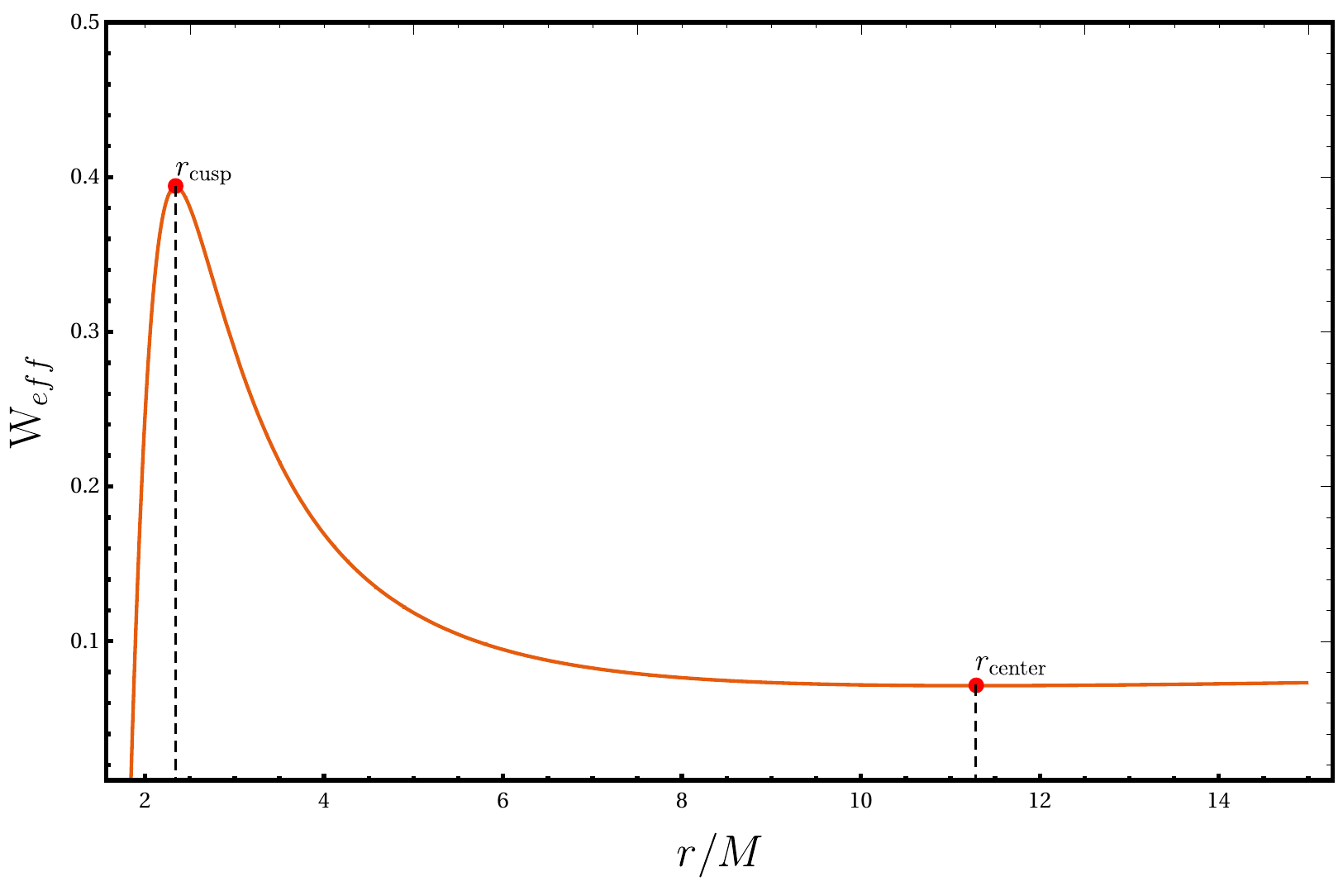} }
\hspace{0.1cm}
\subfigure[] 
{\label{fig6b}\includegraphics[width=8.cm]{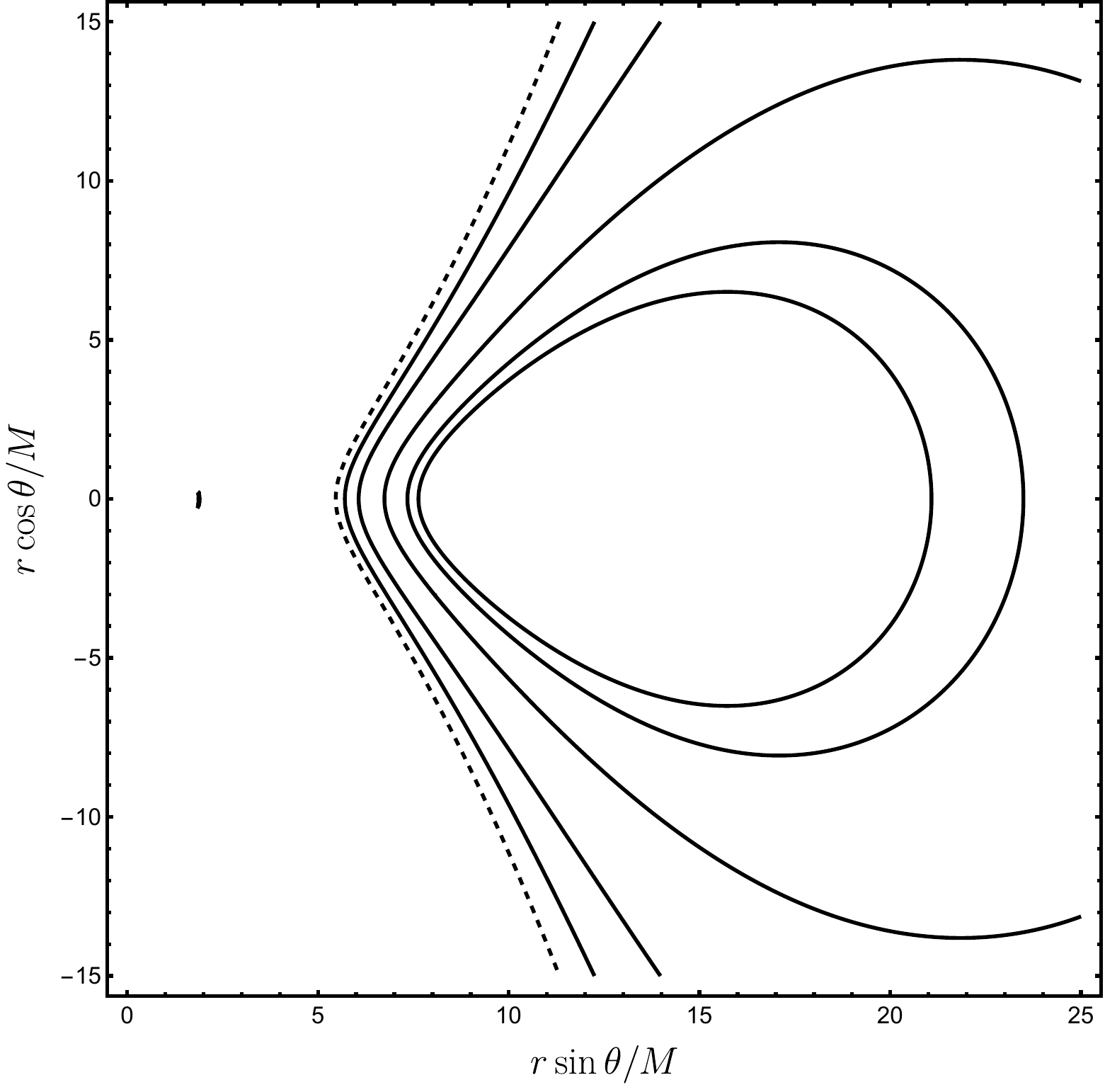} }
\caption{(a): Effective potential \eqref{eq:W2} for  $l=0.189$ and $Q/M=0.5$, with $\ell_0>\ell_{\rm mb}$ and the cusp and torus center identified. (b): Corresponding fluid topology.}\label{Fig6}
\end{figure}

Finally, for $\ell_0>\ell_{\rm mb}$, the potential admits a minimum corresponding to a disk center. In this regime, a local maximum persists at the location previously associated with the cusp, as can be seen in Fig.~\ref{fig6a}, and can therefore be regarded as marginally defined in a geometric sense; however, since it does not satisfy $W_{\rm max}=W(\infty)$ and no self-crossing equipotential surface is formed, it does not correspond to a dynamically active mass-transfer cusp, resulting instead in the open critical configuration displayed here, with a well-defined center. The equipotential surfaces remain closed around the center but open externally at the displayed level; under-filled levels may still define closed finite tori, as can be seen in Fig.~{\color{blue}\ref{fig6b}}. Although Schwarzschild exhibits the most extended structure, as seen in \cite{Font:2002bi}, the inclusion of charge shifts the center inward and increases the radial gradients \cite{PhysRevD.110.124030}. This effect is significantly amplified in the charged KR geometry, where LIV corrections further compress the disk and steepen the intermediate maximum, indicating a more compact equilibrium configuration.  

Unlike the Schwarzschild and RN cases, where the adopted normalization gives $W(\infty)=0$, the charged KR coordinates yield $W(\infty)\neq0$ because the parameter $l$ rescales the temporal metric component. This nonzero asymptotic value is an additive normalization of the potential. Indeed, defining $\bar W\equiv W-W(\infty)$ leaves $\nabla_\mu W$, all critical points, and the equipotential surfaces unchanged. Physical confinement criteria therefore depend on potential differences such as $W-W(\infty)$, rather than on the absolute value of $W(\infty)$ itself; the latter should not, by itself, be interpreted as a dynamical instability.

Table \ref{TAB1} summarizes the representative values used for the Schwarzschild, RN, and charged KR (dubbed here as KR+RN) cases. The rows compare the same qualitative angular-momentum regimes, while the quoted $\ell_0$ values are those adopted for each displayed configuration.
\begin{table*}[ht!]
\centering
\caption{Comparative classification of fluid configurations for Schwarzschild (SCH), RN, and LIV charged RN-like KR (dubbed here as KR+RN) space-times. The KR+RN results correspond to $Q/M=0.5$ and $l=0.189$.}
\renewcommand{\arraystretch}{1.4}
\resizebox{\textwidth}{!}{
\begin{tabular}{|c|c|c|c|c|c|c|c|}
\hline
Regime & Space-time & $\ell_0$& Cusp & Center & Disk type & Equipotential topology & Physical behavior \\
\hline
& SCH &$2.67423$  & No & No & Infinite & Open  & Open, unconfined equilibrium topology \\ 
$\ell_0<\ell_{\rm ms}$ & RN  & $2.5539$ & No & No & Infinite & Open  & Charge-shifted open equilibrium topology \\ 
& KR+RN & $1.8619$ & No & No & Infinite & Open & LIV-shifted open equilibrium topology \\
\hline
& SCH & $3.67423$  & 6 & 6 & Infinite & Critical open surface & Marginal equilibrium boundary \\ 
$\ell_0=\ell_{\rm ms}$ & RN  & $3.5539$ & No & No & None & Open & Marginal equilibrium boundary \\ 
& KR+RN & $2.50657$ & No & No & None & Open & LIV-shifted marginal equilibrium boundary \\
\hline
$3.67423<\ell_0<4$ & SCH & $3.8$ & $4.43$ & $8.8$ & Finite & Closed surfaces & Finite hydrostatic torus \\ 
$3.55394 <\ell_0<3.86851 $ & RN  & $3.7 $ & 4.13 & 8.2 & Finite & Closed surfaces & Finite hydrostatic torus \\ 
$2.50657 < \ell_0 < 2.72719$& KR+RN & $ 2.6$ & 3.18 & 5.99 & Finite & Closed surfaces; reduced radial extent & Narrower stability interval (Lorentz violation effect) \\
\hline
& SCH & $4$ & $4$ & $10.5$ & Infinite & Closed surfaces with cusp; marginally open to infinity & Marginally bound infinite disk \\ 
$\ell_0 = \ell_{\rm mb}$ & RN  & $3.86851$  & $3.7$ & $9.8$ & Infinite &  Closed surfaces with cusp; marginally open to infinity & Marginally bound infinite disc \\ 
& KR+RN & $2.72719$ & $2.81$ & $7.4$ & Infinite & Outer equipotential passes through cusp & Lorentz violation anticipates external unbinding \\
\hline
& SCH &$5$& No & $20.32$ & Infinite & Closed equipotentials without cusp; open externally & Centrifugally supported infinite disk \\ 
$\ell_0 > \ell_{\rm mb}$ & RN  & $4.6851$ & No & $17.50$ & Infinite & Closed equipotentials without cusp; open externally & More compact infinite disk (charge effect) \\ 
& KR+RN & $3.1146$ & No & $14.42$ & Infinite & Open to infinity & Strongly compressed infinite disk \\
\hline
\end{tabular}
}
\label{TAB1}
\end{table*}

\subsection{Direct comparison of the Lorentz violation effect on fluid behavior}

In this section, we provide a direct and visual comparison of the effect of LIV on the fluid behavior. For the neutral KR solution, the characteristic radii are $r_{\rm mb}=4(1-l)M$ and $r_{\rm ms}=6(1-l)M$. We consider $M=1$, and write the corresponding effective angular momenta as
\begin{eqnarray}\label{eq:l_mb_l_ms}
    \ell_{\rm mb}=4 (1-l)^{3/2}\,\,\,\,\, {\rm and}\,\,\,\, \ell_{\rm ms}=3 \sqrt{\frac{3}{2}} (1-l)^{3/2}\,. 
\end{eqnarray}
The $l\rightarrow 0$ limit recovers the Schwarzschild results. To properly visualize the effect of the symmetry-breaking parameter, one must determine the values of $l$ within the interval (\ref{eq:new_KRconstraints}) that yield a value of $\ell_0$ satisfying $\ell_{\rm ms}(l) < \ell_0 < \ell_{\rm mb}(l)$. For example, in the Schwarzschild case (i.e., Eq.(\ref{eq:l_mb_l_ms}) at $l\rightarrow 0$) this restriction corresponds to approximately $3.67<\ell_0<4$. Consequently, one must impose $\ell_{\rm mb}(l) > 3.67$ and $\ell_{\rm ms}(l)<4$, which leads to the following interval
\begin{eqnarray}
    1-\left(\frac{4}{3\sqrt{3/2}}\right)^{2/3}<l< 1-\left(\frac{3.67}{4}\right)^{2/3}\,. 
\end{eqnarray}
Thus, any value of $l$ within this interval returns values of $\ell_{\rm ms}$ and $\ell_{\rm mb}$ that overlap with the respective Schwarzschild range. A similar methodology can be applied to the charged KR solution to determine the interval of $l$ that allows a direct comparison with the fluid behavior in a RN setting. However, due to the length of the resulting expressions, we shall treat this problem numerically instead.

\begin{figure}[t!]
\centering
\includegraphics[width=8.cm]{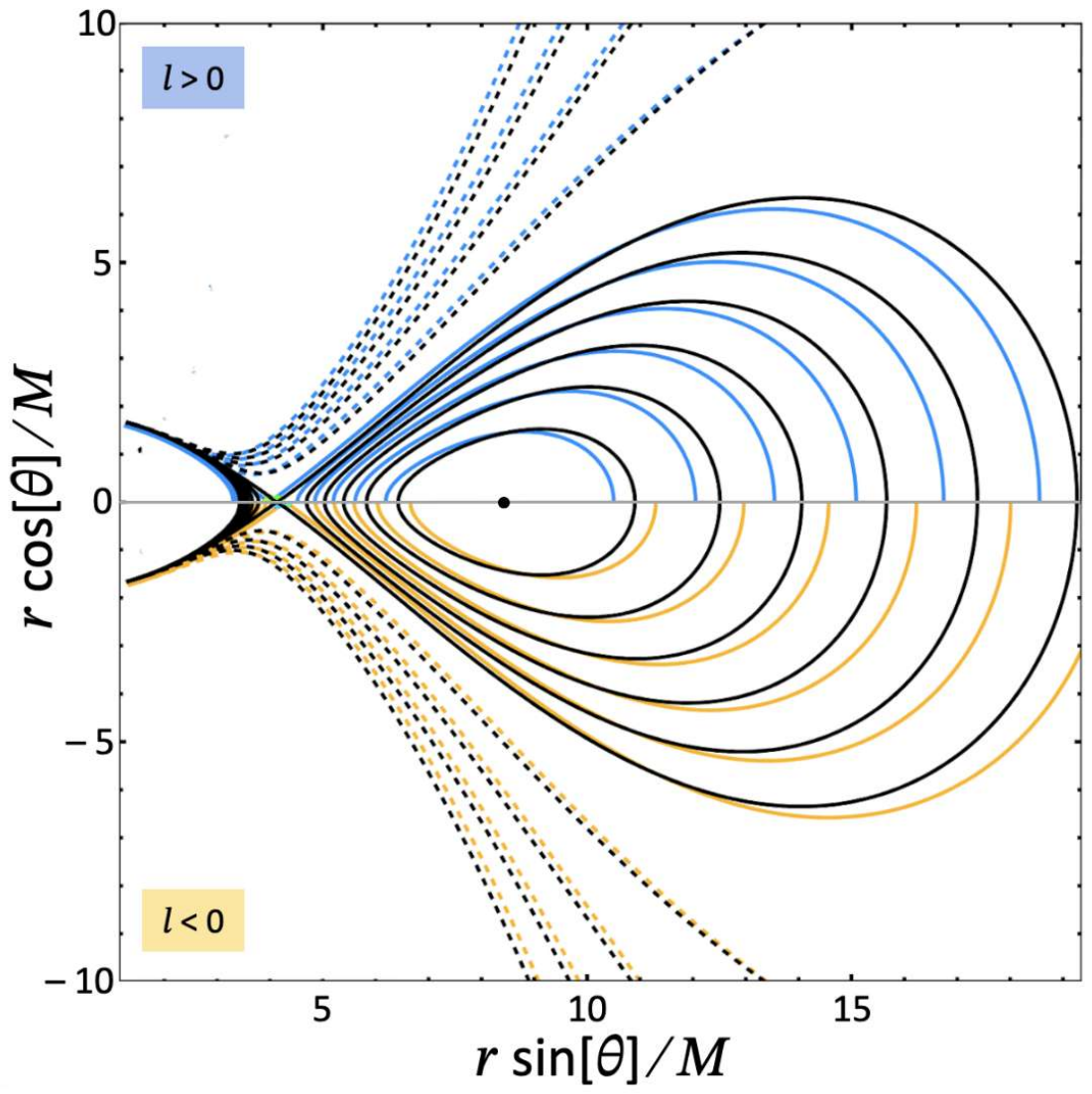} 
\caption{Comparison of the fluid configurations in the RN solution (solid black curve) and the charged KR+RN solution (solid green and blue curves) for the same fluid topology,  with $Q/M=0.5$ and values of  $l=0.03$  and   $l=-0.03$.}
\label{fig:comparison}
\end{figure}

For $Q/M=0.5$, any value of $l$ within the interval $-0.05452425711 < l < 0.05106947343$ produces a fluid topology identical to that of an RN black hole. Fig.\,\ref{fig:comparison} highlights the effect of $l$, which acts as an effective radial rescaling of the fluid equilibrium structure. For $l>0$, both the cusp and the torus center are shifted towards smaller radii, whereas for $l<0$ the shift occurs towards larger radii. The solid green (blue) curves correspond to the charged KR solution, while the solid black curves represent the canonical RN case. The orange and black points indicate, respectively, the cusp and center of the RN torus, whereas the red and brown points mark the corresponding locations for the charged KR solution.

\section{Summary and conclusion}\label{Sec:Conclusion}

In this work we have investigated, within a fully relativistic and self-consistent framework, how Lorentz-violating corrections modify not only the location of characteristic orbits but also the global equilibrium structure of thick accretion disks. To this end we considered a spontaneous symmetry-breaking of Lorentz symmetry given by the Kalb-Ramond field and, more explicitly, a recently found solution of KR gravity with an electromagnetic charge.

We first constrained the parameter space of the charged KR space-time using the shadow radius inferred from the Event Horizon Telescope observations of Sgr A*. The inclusion of the electric charge introduces a degeneracy between $Q$ and $l$, enlarging the region compatible with the observed shadow size and yielding the constraint $-0.3900 \leq l \leq 0.1898$. Such a constraint emerged from the union of the admissible intervals of $l$ as the charge increases. This result shows that, although the shadow provides an important probe of deviations from GR, the simultaneous variation of charge and Lorentz symmetry-breaking effects can significantly modify the observational constraints, making the shadow radius alone insufficient to disentangle the effects of the two parameters. In this regard, observables such as photon ring relative brightness~\cite{daSilva:2023jxa} and photon ring auto-correlations~\cite{Hadar:2020fda} in time-averaged images, might allow one to distinguish between a RN black hole and a charged KR black hole when their shadow sizes are degenerate.

The impact of the LIV corrections coupled with the electric charge becomes more evident when considering the accretion properties because they also strongly inherit the dependence on the black hole parameters. Our findings show that the presence of electric charge increases the emitted flux, temperature, spectral output, torque, and conversion efficiency, while the Lorentz symmetry-breaking parameter produces systematic deviations in all these quantities: positive (negative) values of $l$ enhance (suppress) the radiative output, with these differences becoming more pronounced in the charged configurations. In particular, the high-frequency regime of the emission spectrum is more sensitive to these modifications, leading to the possibility of distinguishing between the presented space-time geometries through astrophysical observations of the emission spectra from accretion discs.

In contrast to previous analyses restricted to Schwarzschild or RN backgrounds \cite{Font:2002bi, PhysRevD.110.124030}, we showed that the parameter $l$ systematically changes the radial extent of the stationary configurations and modifies the interval between $\ell_{\rm ms}$ and $\ell_{\rm mb}$ that supports finite cusp-bearing tori. Outside this interval the equipotential topology becomes open or unconfined. These are statements about stationary equilibrium structure; establishing whether LIV enhances or suppresses a genuine runaway instability requires a time-dependent relativistic hydrodynamical analysis with mass transfer and backreaction.

As shown in Figs.~\ref{fig:ell_eff} and~\ref{fig:ISCO_hor}, the effective angular momentum and the location of the outer horizon depend explicitly on $l$, leading to systematic shifts in the characteristic orbits. In particular, positive values of $l$ make the geometry more compact and bring the orbits closer to the black hole, whereas negative values produce the opposite effect, indicating a strengthening of the effective gravitational potential. Nevertheless, the hierarchy $r_+<r_{\rm mb}<r_{\rm ms}$ is preserved throughout the analyzed parameter range.

The thick disk structure is strongly influenced by these modifications. As illustrated in Figs.~\ref{Fig2}-\ref{Fig6}, the effective potential determines different fluid configuration regimes. Fig.~\ref{Fig2} shows the formation of a finite torus with a well-defined cusp and center for $\ell_{\rm ms}<\ell_0<\ell_{\rm mb}$, characterizing a state of hydrostatic equilibrium. On the other hand, Fig.~\ref{Fig3} indicates that, in the marginally stable case ($\ell_0=\ell_{\rm ms}$), the potential well has already vanished, leaving only an unstable critical point, which marks the corresponding transition in stationary equilibrium topology. Figs.~\ref{Fig4}-\ref{Fig6} further reinforce this behavior, showing that, outside the stability interval, the fluid forms open configurations or infinite disks, without effective confinement.

Lastly, the direct comparison presented in Fig.\,\ref{fig:comparison} shows that the effect of $l$ may be interpreted as a radial rescaling of the fluid structure. For $l>0$, the cusp and torus center are shifted towards smaller radii in the charged KR case relative to RN, indicating a more compact configuration with steeper gradients. In contrast, for $l<0$, the displacement occurs towards larger radii, leading to the opposite radial trend. Thus, the sign of $l$ determines the direction of the radial rescaling induced by LIV effects, consequently modifying the radial extent and equilibrium properties of the disks and the transition between confined and unconfined configurations.

To conclude, our results establish that LIV, through the charged KR geometry, is not a small perturbative effect on fluid configurations, but rather a mechanism capable of reshaping the disk morphology and equilibrium properties, with potential implications for accretion physics and observational signatures in strong gravity regimes.

\section*{Acknowledgements}

FSNL acknowledges support from the Fundação
para a Ciência e a Tecnologia (FCT) Scientific
Employment Stimulus contract with reference
CEECINST/00032/2018, and funding through the
research grants UID/04434/2025. ELBJ (supported by CNPq/PQ 307085/2026-0) and MER thank Conselho
Nacional de Desenvolvimento Científico e Tecnológico
(CNPq), Brazil, for partial financial support. DRG acknowledges financial support by the Spanish National
Grants PID2022-138607NBI00 and CNS2024-154444
grants, funded by MICIU/AEI/10.13039/501100011033 (Spain).
This study was financed in part by the Coordenação de
Aperfeiçoamento de Pessoal de Nível Superior - Brasil
(CAPES) - Finance Code 001.

\bibliography{ref_revised}

@article{LIGOScientific:2016aoc,
    author = "Abbott, B. P. and others",
    collaboration = "LIGO Scientific, Virgo",
    title = "{Observation of Gravitational Waves from a Binary Black Hole Merger}",
    eprint = "1602.03837",
    archivePrefix = "arXiv",
    primaryClass = "gr-qc",
    reportNumber = "LIGO-P150914",
    doi = "10.1103/PhysRevLett.116.061102",
    journal = "Phys. Rev. Lett.",
    volume = "116",
    number = "6",
    pages = "061102",
    year = "2016"
}

@article{LIGOScientific:2016vlm,
    author = "Abbott, B. P. and others",
    collaboration = "LIGO Scientific, Virgo",
    title = "{Properties of the Binary Black Hole Merger GW150914}",
    eprint = "1602.03840",
    archivePrefix = "arXiv",
    primaryClass = "gr-qc",
    reportNumber = "LIGO-P1500218",
    doi = "10.1103/PhysRevLett.116.241102",
    journal = "Phys. Rev. Lett.",
    volume = "116",
    number = "24",
    pages = "241102",
    year = "2016"
}

@article{EventHorizonTelescope:2019dse,
    author = "Akiyama, Kazunori and others",
    collaboration = "Event Horizon Telescope",
    title = "{First M87 Event Horizon Telescope Results. I. The Shadow of the Supermassive Black Hole}",
    eprint = "1906.11238",
    archivePrefix = "arXiv",
    primaryClass = "astro-ph.GA",
    doi = "10.3847/2041-8213/ab0ec7",
    journal = "Astrophys. J. Lett.",
    volume = "875",
    pages = "L1",
    year = "2019"
}

@article{EventHorizonTelescope:2019uob,
    author = "Akiyama, Kazunori and others",
    collaboration = "Event Horizon Telescope",
    title = "{First M87 Event Horizon Telescope Results. II. Array and Instrumentation}",
    eprint = "1906.11239",
    archivePrefix = "arXiv",
    primaryClass = "astro-ph.IM",
    doi = "10.3847/2041-8213/ab0c96",
    journal = "Astrophys. J. Lett.",
    volume = "875",
    number = "1",
    pages = "L2",
    year = "2019"
}

@article{EventHorizonTelescope:2019jan,
    author = "Akiyama, Kazunori and others",
    collaboration = "Event Horizon Telescope",
    title = "{First M87 Event Horizon Telescope Results. III. Data Processing and Calibration}",
    eprint = "1906.11240",
    archivePrefix = "arXiv",
    primaryClass = "astro-ph.GA",
    doi = "10.3847/2041-8213/ab0c57",
    journal = "Astrophys. J. Lett.",
    volume = "875",
    number = "1",
    pages = "L3",
    year = "2019"
}

@article{EventHorizonTelescope:2019ths,
    author = "Akiyama, Kazunori and others",
    collaboration = "Event Horizon Telescope",
    title = "{First M87 Event Horizon Telescope Results. IV. Imaging the Central Supermassive Black Hole}",
    eprint = "1906.11241",
    archivePrefix = "arXiv",
    primaryClass = "astro-ph.GA",
    doi = "10.3847/2041-8213/ab0e85",
    journal = "Astrophys. J. Lett.",
    volume = "875",
    number = "1",
    pages = "L4",
    year = "2019"
}

@article{EventHorizonTelescope:2019pgp,
    author = "Akiyama, Kazunori and others",
    collaboration = "Event Horizon Telescope",
    title = "{First M87 Event Horizon Telescope Results. V. Physical Origin of the Asymmetric Ring}",
    eprint = "1906.11242",
    archivePrefix = "arXiv",
    primaryClass = "astro-ph.GA",
    doi = "10.3847/2041-8213/ab0f43",
    journal = "Astrophys. J. Lett.",
    volume = "875",
    number = "1",
    pages = "L5",
    year = "2019"
}

@article{EventHorizonTelescope:2019ggy,
    author = "Akiyama, Kazunori and others",
    collaboration = "Event Horizon Telescope",
    title = "{First M87 Event Horizon Telescope Results. VI. The Shadow and Mass of the Central Black Hole}",
    eprint = "1906.11243",
    archivePrefix = "arXiv",
    primaryClass = "astro-ph.GA",
    doi = "10.3847/2041-8213/ab1141",
    journal = "Astrophys. J. Lett.",
    volume = "875",
    number = "1",
    pages = "L6",
    year = "2019"
}

@article{EventHorizonTelescope:2022wkp,
    author = "Akiyama, Kazunori and others",
    collaboration = "Event Horizon Telescope",
    title = "{First Sagittarius A* Event Horizon Telescope Results. I. The Shadow of the Supermassive Black Hole in the Center of the Milky Way}",
    eprint = "2311.08680",
    archivePrefix = "arXiv",
    primaryClass = "astro-ph.HE",
    doi = "10.3847/2041-8213/ac6674",
    journal = "Astrophys. J. Lett.",
    volume = "930",
    number = "2",
    pages = "L12",
    year = "2022"
}

@article{EventHorizonTelescope:2022apq,
    author = "Akiyama, Kazunori and others",
    collaboration = "Event Horizon Telescope",
    title = "{First Sagittarius A* Event Horizon Telescope Results. II. EHT and Multiwavelength Observations, Data Processing, and Calibration}",
    eprint = "2311.08679",
    archivePrefix = "arXiv",
    primaryClass = "astro-ph.HE",
    reportNumber = "FERMILAB-PUB-22-418-PPD",
    doi = "10.3847/2041-8213/ac6675",
    journal = "Astrophys. J. Lett.",
    volume = "930",
    number = "2",
    pages = "L13",
    year = "2022"
}

@article{EventHorizonTelescope:2022wok,
    author = "Akiyama, Kazunori and others",
    collaboration = "Event Horizon Telescope",
    title = "{First Sagittarius A* Event Horizon Telescope Results. III. Imaging of the Galactic Center Supermassive Black Hole}",
    eprint = "2311.09479",
    archivePrefix = "arXiv",
    primaryClass = "astro-ph.HE",
    doi = "10.3847/2041-8213/ac6429",
    journal = "Astrophys. J. Lett.",
    volume = "930",
    number = "2",
    pages = "L14",
    year = "2022"
}

@article{EventHorizonTelescope:2022exc,
    author = "Akiyama, Kazunori and others",
    collaboration = "Event Horizon Telescope",
    title = "{First Sagittarius A* Event Horizon Telescope Results. IV. Variability, Morphology, and Black Hole Mass}",
    eprint = "2311.08697",
    archivePrefix = "arXiv",
    primaryClass = "astro-ph.HE",
    reportNumber = "FERMILAB-PUB-22-423-PPD",
    doi = "10.3847/2041-8213/ac6736",
    journal = "Astrophys. J. Lett.",
    volume = "930",
    number = "2",
    pages = "L15",
    year = "2022"
}

@article{EventHorizonTelescope:2022urf,
    author = "Akiyama, Kazunori and others",
    collaboration = "Event Horizon Telescope",
    title = "{First Sagittarius A* Event Horizon Telescope Results. V. Testing Astrophysical Models of the Galactic Center Black Hole}",
    eprint = "2311.09478",
    archivePrefix = "arXiv",
    primaryClass = "astro-ph.HE",
    reportNumber = "FERMILAB-PUB-22-419-PPD",
    doi = "10.3847/2041-8213/ac6672",
    journal = "Astrophys. J. Lett.",
    volume = "930",
    number = "2",
    pages = "L16",
    year = "2022"
}

@article{EventHorizonTelescope:2022xqj,
    author = "Akiyama, Kazunori and others",
    collaboration = "Event Horizon Telescope",
    title = "{First Sagittarius A* Event Horizon Telescope Results. VI. Testing the Black Hole Metric}",
    eprint = "2311.09484",
    archivePrefix = "arXiv",
    primaryClass = "astro-ph.HE",
    reportNumber = "FERMILAB-PUB-22-422-PPD",
    doi = "10.3847/2041-8213/ac6756",
    journal = "Astrophys. J. Lett.",
    volume = "930",
    number = "2",
    pages = "L17",
    year = "2022"
}

@article{Vagnozzi:2022moj,
    author = "Vagnozzi, Sunny and others",
    title = "{Horizon-scale tests of gravity theories and fundamental physics from the Event Horizon Telescope image of Sagittarius A}",
    eprint = "2205.07787",
    archivePrefix = "arXiv",
    primaryClass = "gr-qc",
    reportNumber = "UCI-HEP-TR-2022-07",
    doi = "10.1088/1361-6382/acd97b",
    journal = "Class. Quant. Grav.",
    volume = "40",
    number = "16",
    pages = "165007",
    year = "2023"
}

@article{bondi1952spherically,
  title={On spherically symmetrical accretion},
  author={Bondi, Hermann},
  journal={Monthly Notices of the Royal Astronomical Society},
  volume={112},
  number={2},
  pages={195--204},
  year={1952},
  publisher={Oxford University Press Oxford, UK}
}

@article{Armitage:2020owb,
    author = "Armitage, Philip J.",
    title = "{Bondi on spherically symmetric accretion}",
    eprint = "2004.00203",
    archivePrefix = "arXiv",
    primaryClass = "astro-ph.HE",
    doi = "10.1093/astrogeo/ataa032",
    journal = "Astron. Geophys.",
    volume = "61",
    number = "2",
    pages = "2.40--2.42",
    year = "2020"
}

@article{Michel1972,
  author       = {F. Curtis Michel},
  title        = {Accretion of Matter by Condensed Objects},
  journal      = {Astrophysics and Space Science},
  year         = {1972},
  volume       = {15},
  number       = {1},
  pages        = {153--160},
  doi          = {10.1007/BF00649949},
  publisher    = {Springer}
}

@article{Moncrief1980,
  author       = {Vincent Moncrief},
  title        = {Stability of Stationary, Spherical Accretion onto a Schwarzschild Black Hole},
  journal      = {The Astrophysical Journal},
  year         = {1980},
  volume       = {235},
  pages        = {1038--1046},
  doi          = {10.1086/157707}
}

@article{Babichev:2004yx,
    author = "Babichev, E. and Dokuchaev, V. and Eroshenko, Yu.",
    title = "{Black hole mass decreasing due to phantom energy accretion}",
    eprint = "gr-qc/0402089",
    archivePrefix = "arXiv",
    doi = "10.1103/PhysRevLett.93.021102",
    journal = "Phys. Rev. Lett.",
    volume = "93",
    pages = "021102",
    year = "2004"
}

@article{Babichev:2013vji,
    author = "Babichev, E. O. and Dokuchaev, V. I. and Eroshenko, Yu N.",
    title = "{Black holes in the presence of dark energy}",
    eprint = "1406.0841",
    archivePrefix = "arXiv",
    primaryClass = "gr-qc",
    reportNumber = "LPT-ORSAY-14-33",
    doi = "10.3367/UFNe.0183.201312a.1257",
    journal = "Phys. Usp.",
    volume = "56",
    pages = "1155--1175",
    year = "2013"
}

@article{Jamil:2010skm,
    author = "Jamil, Mubasher and Qadir, Asghar",
    title = "{Primordial Black Holes in Phantom Cosmology}",
    eprint = "0908.0444",
    archivePrefix = "arXiv",
    primaryClass = "gr-qc",
    doi = "10.1007/s10714-010-0928-1",
    journal = "Gen. Rel. Grav.",
    volume = "43",
    pages = "1069--1082",
    year = "2011"
}

@article{novikov1973astrophysics,
  title={Astrophysics of black holes},
  author={Novikov, Igor D and Thorne, Kip S},
  journal={Black holes (Les astres occlus)},
  volume={1},
  pages={343--450},
  year={1973}
}

@article{page1974disk,
  title={Disk-accretion onto a black hole. Time-averaged structure of accretion disk},
  author={Page, Don N and Thorne, Kip S},
  journal={Astrophysical Journal, Vol. 191, pp. 499-506 (1974)},
  volume={191},
  pages={499--506},
  year={1974}
}

@article{Lahiri:2020sza,
    author = "Lahiri, Sayantani and Gimeno-Soler, Sergio and Font, Jos{\'e} A. and Mej{\'\i}as, Alejandro Mus",
    title = "{Stationary models of magnetized viscous tori around a Schwarzschild black hole}",
    eprint = "2012.06835",
    archivePrefix = "arXiv",
    primaryClass = "gr-qc",
    doi = "10.1103/PhysRevD.103.044034",
    journal = "Phys. Rev. D",
    volume = "103",
    number = "4",
    pages = "044034",
    year = "2021"
}

@article{Chen:2024jsv,
    author = "Chen, Chengjia and Pan, Qiyuan and Jing, Jiliang",
    title = "{Geometrically thick equilibrium tori around a Schwarzschild black hole in swirling universes}",
    eprint = "2402.02789",
    archivePrefix = "arXiv",
    primaryClass = "gr-qc",
    doi = "10.1140/epjc/s10052-024-13415-z",
    journal = "Eur. Phys. J. C",
    volume = "84",
    number = "10",
    pages = "1040",
    year = "2024"
}

@article{Ricarte:2022gcl,
    author = "Ricarte, Angelo and Gammie, Charles and Narayan, Ramesh and Prather, Ben S.",
    title = "{Probing plasma physics with spectral index maps of accreting black holes on event horizon scales}",
    eprint = "2202.02408",
    archivePrefix = "arXiv",
    primaryClass = "astro-ph.HE",
    doi = "10.1093/mnras/stac3796",
    journal = "Mon. Not. Roy. Astron. Soc.",
    volume = "519",
    number = "3",
    pages = "4203--4220",
    year = "2023"
}

@article{Dhruv:2024igk,
    author = "Dhruv, Vedant and Prather, Ben and Wong, George N. and Gammie, Charles F.",
    title = "{A Survey of General Relativistic Magnetohydrodynamic Models for Black Hole Accretion Systems}",
    eprint = "2411.12647",
    archivePrefix = "arXiv",
    primaryClass = "astro-ph.HE",
    doi = "10.3847/1538-4365/adaea6",
    journal = "Astrophys. J. Suppl.",
    volume = "277",
    number = "1",
    pages = "16",
    year = "2025"
}

@article{Abramowicz:2011xu,
    author = "Abramowicz, Marek A. and Fragile, P. Chris",
    title = "{Foundations of Black Hole Accretion Disk Theory}",
    eprint = "1104.5499",
    archivePrefix = "arXiv",
    primaryClass = "astro-ph.HE",
    reportNumber = "NSF-KITP-12-055",
    doi = "10.12942/lrr-2013-1",
    journal = "Living Rev. Rel.",
    volume = "16",
    pages = "1",
    year = "2013"
}

@article{Font:2002bi,
    author = "Font, Jose A. and Daigne, Frederic",
    title = "{The Runaway instability of thick discs around black holes. 1. The Constant angular momentum case}",
    eprint = "astro-ph/0203403",
    archivePrefix = "arXiv",
    doi = "10.1046/j.1365-8711.2002.05515.x",
    journal = "Mon. Not. Roy. Astron. Soc.",
    volume = "334",
    pages = "383",
    year = "2002"
}

@article{Rezzolla:2003re,
    author = "Rezzolla, Luciano and Zanotti, Olindo and Font, Jose A.",
    title = "{Dynamics of thick discs around Schwarzschild-de Sitter black holes}",
    eprint = "gr-qc/0310045",
    archivePrefix = "arXiv",
    doi = "10.1051/0004-6361:20031457",
    journal = "Astron. Astrophys.",
    volume = "412",
    pages = "603",
    year = "2003"
}

@article{PhysRevD.110.124030,
  title = {Fluid figures of equilibrium orbiting Reissner-Nordstr\"om black holes and naked singularities},
  author = {Mishra, Ruchi and Krajewski, Tomasz and Klu\ifmmode \acute{z}\else \'{z}\fi{}niak, W\l{}odek},
  journal = {Phys. Rev. D},
  volume = {110},
  issue = {12},
  pages = {124030},
  numpages = {14},
  year = {2024},
  month = {Dec},
  publisher = {American Physical Society},
  doi = {10.1103/PhysRevD.110.124030},
  url = {https://link.aps.org/doi/10.1103/PhysRevD.110.124030}
}

@article{Senovilla:2014gza,
    author = "Senovilla, Jos{\'e} M. M. and Garfinkle, David",
    title = "{The 1965 Penrose singularity theorem}",
    eprint = "1410.5226",
    archivePrefix = "arXiv",
    primaryClass = "gr-qc",
    doi = "10.1088/0264-9381/32/12/124008",
    journal = "Class. Quant. Grav.",
    volume = "32",
    number = "12",
    pages = "124008",
    year = "2015"
}

@article{Carlip:2001wq,
    author = "Carlip, Steven",
    title = "{Quantum gravity: A Progress report}",
    eprint = "gr-qc/0108040",
    archivePrefix = "arXiv",
    reportNumber = "UCD-2001-04",
    doi = "10.1088/0034-4885/64/8/301",
    journal = "Rept. Prog. Phys.",
    volume = "64",
    pages = "885",
    year = "2001"
}

@article{Kostelecky:1988zi,
    author = "Kostelecky, V. Alan and Samuel, Stuart",
    title = "{Spontaneous Breaking of Lorentz Symmetry in String Theory}",
    reportNumber = "IUHET-139, CCNY-HEP-88/4",
    doi = "10.1103/PhysRevD.39.683",
    journal = "Phys. Rev. D",
    volume = "39",
    pages = "683",
    year = "1989"
}

@article{PhysRevLett.63.224,
  title = {Phenomenological gravitational constraints on strings and higher-dimensional theories},
  author = {Kosteleck\'y, V. Alan and Samuel, Stuart},
  journal = {Phys. Rev. Lett.},
  volume = {63},
  issue = {3},
  pages = {224--227},
  numpages = {0},
  year = {1989},
  month = {Jul},
  publisher = {American Physical Society},
  doi = {10.1103/PhysRevLett.63.224},
  url = {https://link.aps.org/doi/10.1103/PhysRevLett.63.224}
}

@article{PhysRevD.40.1886,
  title = {Gravitational phenomenology in higher-dimensional theories and strings},
  author = {Kosteleck\'y, V. Alan and Samuel, Stuart},
  journal = {Phys. Rev. D},
  volume = {40},
  issue = {6},
  pages = {1886--1903},
  numpages = {0},
  year = {1989},
  month = {Sep},
  publisher = {American Physical Society},
  doi = {10.1103/PhysRevD.40.1886},
  url = {https://link.aps.org/doi/10.1103/PhysRevD.40.1886}
}

@article{Colladay:1996iz,
    author = "Colladay, Don and Kostelecky, V. Alan",
    title = "{CPT violation and the standard model}",
    eprint = "hep-ph/9703464",
    archivePrefix = "arXiv",
    reportNumber = "IUHET-354",
    doi = "10.1103/PhysRevD.55.6760",
    journal = "Phys. Rev. D",
    volume = "55",
    pages = "6760--6774",
    year = "1997"
}

@book{Rovelli2004,
  author    = {Carlo Rovelli},
  title     = {Quantum Gravity},
  publisher = {Cambridge University Press},
  address   = {Cambridge},
  year      = {2004},
  doi       = {10.1017/CBO9780511755804},
  isbn      = {9780521837330}
}

@article{Ashtekar:2004eh,
    author = "Ashtekar, Abhay and Lewandowski, Jerzy",
    title = "{Background independent quantum gravity: A Status report}",
    eprint = "gr-qc/0404018",
    archivePrefix = "arXiv",
    doi = "10.1088/0264-9381/21/15/R01",
    journal = "Class. Quant. Grav.",
    volume = "21",
    pages = "R53",
    year = "2004"
}

@article{Gambini:1998it,
    author = "Gambini, Rodolfo and Pullin, Jorge",
    title = "{Nonstandard optics from quantum space-time}",
    eprint = "gr-qc/9809038",
    archivePrefix = "arXiv",
    reportNumber = "CGPG-98-9-1",
    doi = "10.1103/PhysRevD.59.124021",
    journal = "Phys. Rev. D",
    volume = "59",
    pages = "124021",
    year = "1999"
}

@article{Ellis:1999uh,
    author = "Ellis, John R. and Mavromatos, N. E. and Nanopoulos, Dimitri V.",
    title = "{Quantum gravitational diffusion and stochastic fluctuations in the velocity of light}",
    eprint = "gr-qc/9904068",
    archivePrefix = "arXiv",
    reportNumber = "ACT-3-99, CTP-TAMU-18-99, OUTP-99-05-P",
    doi = "10.1023/A:1001852601248",
    journal = "Gen. Rel. Grav.",
    volume = "32",
    pages = "127--144",
    year = "2000"
}

@article{Douglas:2001ba,
    author = "Douglas, Michael R. and Nekrasov, Nikita A.",
    title = "{Noncommutative field theory}",
    eprint = "hep-th/0106048",
    archivePrefix = "arXiv",
    reportNumber = "ITEP-TH-31-01, IHES-P-01-27, RUNHETC-2001-18",
    doi = "10.1103/RevModPhys.73.977",
    journal = "Rev. Mod. Phys.",
    volume = "73",
    pages = "977--1029",
    year = "2001"
}

@article{Carroll:2001ws,
    author = "Carroll, Sean M. and Harvey, Jeffrey A. and Kostelecky, V. Alan and Lane, Charles D. and Okamoto, Takemi",
    title = "{Noncommutative field theory and Lorentz violation}",
    eprint = "hep-th/0105082",
    archivePrefix = "arXiv",
    reportNumber = "EFI-01-12, IUHET-433",
    doi = "10.1103/PhysRevLett.87.141601",
    journal = "Phys. Rev. Lett.",
    volume = "87",
    pages = "141601",
    year = "2001"
}

@article{Szabo:2001kg,
    author = "Szabo, Richard J.",
    title = "{Quantum field theory on noncommutative spaces}",
    eprint = "hep-th/0109162",
    archivePrefix = "arXiv",
    reportNumber = "HWM-01-35, EMPG-01-14",
    doi = "10.1016/S0370-1573(03)00059-0",
    journal = "Phys. Rept.",
    volume = "378",
    pages = "207--299",
    year = "2003"
}

@article{deRham:2014zqa,
    author = "de Rham, Claudia",
    title = "{Massive Gravity}",
    eprint = "1401.4173",
    archivePrefix = "arXiv",
    primaryClass = "hep-th",
    doi = "10.12942/lrr-2014-7",
    journal = "Living Rev. Rel.",
    volume = "17",
    pages = "7",
    year = "2014"
}

@article{Rham:2015mxa,
    author = "Rham, Claudia",
    editor = "Papantonopoulos, Eleftherios",
    title = "{Introduction to Massive Gravity}",
    doi = "10.1007/978-3-319-10070-8_5",
    journal = "Lect. Notes Phys.",
    volume = "892",
    pages = "139--159",
    year = "2015"
}

@article{Horava:2009uw,
    author = "Horava, Petr",
    title = "{Quantum Gravity at a Lifshitz Point}",
    eprint = "0901.3775",
    archivePrefix = "arXiv",
    primaryClass = "hep-th",
    doi = "10.1103/PhysRevD.79.084008",
    journal = "Phys. Rev. D",
    volume = "79",
    pages = "084008",
    year = "2009"
}

@article{Addazi:2021xuf,
    author = "Addazi, A. and others",
    title = "{Quantum gravity phenomenology at the dawn of the multi-messenger era{\textemdash}A review}",
    eprint = "2111.05659",
    archivePrefix = "arXiv",
    primaryClass = "hep-ph",
    doi = "10.1016/j.ppnp.2022.103948",
    journal = "Prog. Part. Nucl. Phys.",
    volume = "125",
    pages = "103948",
    year = "2022"
}

@article{Kostelecky:2003fs,
    author = "Kostelecky, V. Alan",
    title = "{Gravity, Lorentz violation, and the standard model}",
    eprint = "hep-th/0312310",
    archivePrefix = "arXiv",
    reportNumber = "IUHET-461",
    doi = "10.1103/PhysRevD.69.105009",
    journal = "Phys. Rev. D",
    volume = "69",
    pages = "105009",
    year = "2004"
}

@article{Kostelecky:1991ak,
    author = "Kostelecky, V. Alan and Potting, Robertus",
    title = "{CPT and strings}",
    reportNumber = "IUHET-197",
    doi = "10.1016/0550-3213(91)90071-5",
    journal = "Nucl. Phys. B",
    volume = "359",
    pages = "545--570",
    year = "1991"
}

@article{PhysRevLett.66.1811,
  title = {Photon and graviton masses in string theories},
  author = {Kosteleck\'y, V. Alan and Samuel, Stuart},
  journal = {Phys. Rev. Lett.},
  volume = {66},
  issue = {14},
  pages = {1811--1814},
  numpages = {0},
  year = {1991},
  month = {Apr},
  publisher = {American Physical Society},
  doi = {10.1103/PhysRevLett.66.1811},
  url = {https://link.aps.org/doi/10.1103/PhysRevLett.66.1811}
}

@article{Kostelecky:1995qk,
    author = "Kostelecky, V. Alan and Potting, R.",
    title = "{Expectation values, Lorentz invariance, and CPT in the open bosonic string}",
    eprint = "hep-th/9605088",
    archivePrefix = "arXiv",
    reportNumber = "IUHET-319, UATP-95-05",
    doi = "10.1016/0370-2693(96)00589-8",
    journal = "Phys. Lett. B",
    volume = "381",
    pages = "89--96",
    year = "1996"
}

@article{Kostelecky:2000hz,
    author = "Kostelecky, V. Alan and Potting, Robertus",
    title = "{Analytical construction of a nonperturbative vacuum for the open bosonic string}",
    eprint = "hep-th/0008252",
    archivePrefix = "arXiv",
    reportNumber = "IUHET-426",
    doi = "10.1103/PhysRevD.63.046007",
    journal = "Phys. Rev. D",
    volume = "63",
    pages = "046007",
    year = "2001"
}

@article{Myers:2003fd,
    author = "Myers, Robert C. and Pospelov, Maxim",
    title = "{Ultraviolet modifications of dispersion relations in effective field theory}",
    eprint = "hep-ph/0301124",
    archivePrefix = "arXiv",
    reportNumber = "UVIC-TH-03-01, SUSX-TH-03-001",
    doi = "10.1103/PhysRevLett.90.211601",
    journal = "Phys. Rev. Lett.",
    volume = "90",
    pages = "211601",
    year = "2003"
}

@article{Bluhm:2008yt,
    author = "Bluhm, Robert and Gagne, Nolan L. and Potting, Robertus and Vrublevskis, Arturs",
    title = "{Constraints and Stability in Vector Theories with Spontaneous Lorentz Violation}",
    eprint = "0802.4071",
    archivePrefix = "arXiv",
    primaryClass = "hep-th",
    doi = "10.1103/PhysRevD.77.125007",
    journal = "Phys. Rev. D",
    volume = "77",
    pages = "125007",
    year = "2008",
    note = "[Erratum: Phys.Rev.D 79, 029902 (2009)]"
}

@article{Casana:2017jkc,
    author = "Casana, R. and Cavalcante, A. and Poulis, F. P. and Santos, E. B.",
    title = "{Exact Schwarzschild-like solution in a bumblebee gravity model}",
    eprint = "1711.02273",
    archivePrefix = "arXiv",
    primaryClass = "gr-qc",
    doi = "10.1103/PhysRevD.97.104001",
    journal = "Phys. Rev. D",
    volume = "97",
    number = "10",
    pages = "104001",
    year = "2018"
}

@article{Kalb:1974yc,
    author = "Kalb, Michael and Ramond, Pierre",
    title = "{Classical direct interstring action}",
    doi = "10.1103/PhysRevD.9.2273",
    journal = "Phys. Rev. D",
    volume = "9",
    pages = "2273--2284",
    year = "1974"
}

@article{Altschul:2009ae,
    author = "Altschul, Brett and Bailey, Quentin G. and Kostelecky, V. Alan",
    title = "{Lorentz violation with an antisymmetric tensor}",
    eprint = "0912.4852",
    archivePrefix = "arXiv",
    primaryClass = "gr-qc",
    reportNumber = "IUHET-537",
    doi = "10.1103/PhysRevD.81.065028",
    journal = "Phys. Rev. D",
    volume = "81",
    pages = "065028",
    year = "2010"
}

@article{Aashish:2019ykb,
    author = "Aashish, Sandeep and Panda, Sukanta",
    title = "{Quantum aspects of antisymmetric tensor field with spontaneous Lorentz violation}",
    eprint = "1903.11364",
    archivePrefix = "arXiv",
    primaryClass = "gr-qc",
    doi = "10.1103/PhysRevD.100.065010",
    journal = "Phys. Rev. D",
    volume = "100",
    number = "6",
    pages = "065010",
    year = "2019"
}

@article{Duan:2023gng,
    author = "Duan, Zheng-Qiao and Zhao, Ju-Ying and Yang, Ke",
    title = "{Electrically charged black holes in gravity with a background Kalb{\textendash}Ramond field}",
    eprint = "2310.13555",
    archivePrefix = "arXiv",
    primaryClass = "gr-qc",
    doi = "10.1140/epjc/s10052-024-13188-5",
    journal = "Eur. Phys. J. C",
    volume = "84",
    number = "8",
    pages = "798",
    year = "2024"
}

@article{Lessa:2019bgi,
    author = "Lessa, L. A. and Silva, J. E. G. and Maluf, R. V. and Almeida, C. A. S.",
    title = "{Modified black hole solution with a background Kalb{\textendash}Ramond field}",
    eprint = "1911.10296",
    archivePrefix = "arXiv",
    primaryClass = "gr-qc",
    doi = "10.1140/epjc/s10052-020-7902-1",
    journal = "Eur. Phys. J. C",
    volume = "80",
    number = "4",
    pages = "335",
    year = "2020"
}

@article{Junior:2024ety,
    author = "Junior, Ednaldo L. B. and Junior, Jos{\'e} Tarciso S. S. and Lobo, Francisco S. N. and Rodrigues, Manuel E. and Rubiera-Garcia, Diego and da Silva, Lu{\'\i}s F. Dias and Vieira, Henrique A.",
    title = "{Spontaneous Lorentz symmetry-breaking constraints in Kalb{\textendash}Ramond gravity}",
    eprint = "2405.03291",
    archivePrefix = "arXiv",
    primaryClass = "gr-qc",
    doi = "10.1140/epjc/s10052-024-13619-3",
    journal = "Eur. Phys. J. C",
    volume = "84",
    number = "12",
    pages = "1257",
    year = "2024"
}

@article{Rosa:2023qcv,
    author = "Rosa, Jo{\~a}o Lu{\'\i}s and Macedo, Caio F. B. and Rubiera-Garcia, Diego",
    title = "{Imaging compact boson stars with hot spots and thin accretion disks}",
    eprint = "2303.17296",
    archivePrefix = "arXiv",
    primaryClass = "gr-qc",
    doi = "10.1103/PhysRevD.108.044021",
    journal = "Phys. Rev. D",
    volume = "108",
    number = "4",
    pages = "044021",
    year = "2023"
}

@article{Cunha:2018acu,
    author = "Cunha, Pedro V. P. and Herdeiro, Carlos A. R.",
    title = "{Shadows and strong gravitational lensing: a brief review}",
    eprint = "1801.00860",
    archivePrefix = "arXiv",
    primaryClass = "gr-qc",
    doi = "10.1007/s10714-018-2361-9",
    journal = "Gen. Rel. Grav.",
    volume = "50",
    number = "4",
    pages = "42",
    year = "2018"
}

@article{Gralla:2019xty,
    author = "Gralla, Samuel E. and Holz, Daniel E. and Wald, Robert M.",
    title = "{Black Hole Shadows, Photon Rings, and Lensing Rings}",
    eprint = "1906.00873",
    archivePrefix = "arXiv",
    primaryClass = "astro-ph.HE",
    doi = "10.1103/PhysRevD.100.024018",
    journal = "Phys. Rev. D",
    volume = "100",
    number = "2",
    pages = "024018",
    year = "2019"
}

@article{Chael:2021rjo,
    author = "Chael, Andrew and Johnson, Michael D. and Lupsasca, Alexandru",
    title = "{Observing the Inner Shadow of a Black Hole: A Direct View of the Event Horizon}",
    eprint = "2106.00683",
    archivePrefix = "arXiv",
    primaryClass = "astro-ph.HE",
    doi = "10.3847/1538-4357/ac09ee",
    journal = "Astrophys. J.",
    volume = "918",
    number = "1",
    pages = "6",
    year = "2021"
}

@article{perlick2022,
  title={Calculating black hole shadows: Review of analytical studies},
  author={Perlick, Volker and Tsupko, Oleg Yu},
  journal={Physics Reports},
  volume={947},
  pages={1--39},
  year={2022},
  publisher={Elsevier}
}

@article{claudel2001,
  title={The geometry of photon surfaces},
  author={Claudel, Clarissa-Marie and Virbhadra, Kumar Shwetketu and Ellis, George FR},
  journal={Journal of Mathematical Physics},
  volume={42},
  number={2},
  pages={818--838},
  year={2001},
  publisher={American Institute of Physics}
}

@article{Do:2019txf,
    author = "Do, Tuan and others",
    title = "{Relativistic redshift of the star S0-2 orbiting the Galactic center supermassive black hole}",
    eprint = "1907.10731",
    archivePrefix = "arXiv",
    primaryClass = "astro-ph.GA",
    doi = "10.1126/science.aav8137",
    journal = "Science",
    volume = "365",
    number = "6454",
    pages = "664--668",
    year = "2019"
}

@article{GRAVITY:2020gka,
    author = "Abuter, R. and others",
    collaboration = "GRAVITY",
    title = "{Detection of the Schwarzschild precession in the orbit of the star S2 near the Galactic centre massive black hole}",
    eprint = "2004.07187",
    archivePrefix = "arXiv",
    primaryClass = "astro-ph.GA",
    doi = "10.1051/0004-6361/202037813",
    journal = "Astron. Astrophys.",
    volume = "636",
    pages = "L5",
    year = "2020"
}

@article{Torres:2002td,
    author = "Torres, Diego F.",
    title = "{Accretion disc onto a static nonbaryonic compact object}",
    eprint = "hep-ph/0201154",
    archivePrefix = "arXiv",
    doi = "10.1016/S0550-3213(02)00038-X",
    journal = "Nucl. Phys. B",
    volume = "626",
    pages = "377--394",
    year = "2002"
}

@article{Abramowicz1974,
  author  = {M. A. Abramowicz},
  title   = {Theory of Level Surfaces Inside Relativistic, Rotating Stars. II},
  journal = {Acta Astronomica},
  volume  = {24},
  pages   = {45--53},
  year    = {1974}
}

@article{PaczynskiAbramowicz1982,
  author  = {B. Paczynski and M. A. Abramowicz},
  title   = {A Model of a Thick Disk with Equatorial Accretion},
  journal = {The Astrophysical Journal},
  volume  = {253},
  pages   = {897--907},
  year    = {1982},
  doi     = {10.1086/159689}
}

@article{Boyer1965,
  author  = {R. H. Boyer},
  title   = {Rotating Fluid Masses in General Relativity},
  journal = {Proceedings of the Cambridge Philosophical Society},
  volume  = {61},
  pages   = {527--530},
  year    = {1965}
}

@article{daSilva:2023jxa,
    author = "da Silva, Lu{\'\i}s F. Dias and Lobo, Francisco S. N. and Olmo, Gonzalo J. and Rubiera-Garcia, Diego",
    title = "{Photon rings as tests for alternative spherically symmetric geometries with thin accretion disks}",
    eprint = "2307.06778",
    archivePrefix = "arXiv",
    primaryClass = "gr-qc",
    doi = "10.1103/PhysRevD.108.084055",
    journal = "Phys. Rev. D",
    volume = "108",
    number = "8",
    pages = "084055",
    year = "2023"
}

@article{Hadar:2020fda,
    author = "Hadar, Shahar and Johnson, Michael D. and Lupsasca, Alexandru and Wong, George N.",
    title = "{Photon Ring Autocorrelations}",
    eprint = "2010.03683",
    archivePrefix = "arXiv",
    primaryClass = "gr-qc",
    doi = "10.1103/PhysRevD.103.104038",
    journal = "Phys. Rev. D",
    volume = "103",
    number = "10",
    pages = "104038",
    year = "2021"
}

\end{document}